\documentclass{aa}

\usepackage{graphicx}
\usepackage{txfonts}
\usepackage{hyperref}
\usepackage{siunitx}
\usepackage{natbib}
\usepackage{lineno}
\usepackage{color}

\DeclareMathOperator\erf{erf}

\def\reff{R_{\mathrm{e}}}

\def\Sref#1{Section~\ref{#1}}
\def\Fref#1{Figure~\ref{#1}}

\newcommand{\gmr}{$\mathit{\Gamma_{*,10}} - M_{*,10}$ relation}

\begin{document}

   \title{The $10~\rm{kpc}$ collar of early-type galaxies --- probing evolution by focusing on the inner stellar density profile}
   \titlerunning{The $10~\rm{kpc}$ collar}
   \authorrunning{Rongfu et al.}

   \author{Rongfu Liu\inst{\ref{sjtu1}\fnmsep\thanks{\email{liurongfu@sjtu.edu.cn} }} \and
          Alessandro Sonnenfeld\inst{\ref{sjtu1},\ref{sjtu2},\ref{sjtu3}\fnmsep\thanks{\email{sonnenfeld@sjtu.edu.cn} }}\and 
          Carlo Nipoti\inst{\ref{Italy}} \and
          Rui Li\inst{\ref{Zhengzhou}} 
          }

   \institute{
Department of Astronomy, School of Physics and Astronomy, Shanghai Jiao Tong University, Shanghai 200240, China\label{sjtu1} \and
Shanghai Key Laboratory for Particle Physics and Cosmology, Shanghai Jiao Tong University, Shanghai 200240, China\label{sjtu2} \and
Key Laboratory for Particle Physics, Astrophysics and Cosmology, Ministry of Education, Shanghai Jiao Tong University, Shanghai 200240, China\label{sjtu3} \and
Dipartimento di Fisica e Astronomia “Augusto Righi”, Alma Mater Studiorum - Università di Bologna, via Gobetti 93/2, 40129, Bologna, Italy\label{Italy} \and
Institute for Astrophysics, School of Physics, Zhengzhou University, Zhengzhou 450001, China\label{Zhengzhou}
}

   \date{}
    \abstract
    {
        The post-quenching evolution process of early-type galaxies (ETGs), which is typically driven by mergers, is still not fully understood. 
The amount of growth in stellar mass and size incurred after quenching is still under debate.
}
   {
In this work we aim to investigate the late evolution of ETGs, both observationally and theoretically, by focusing on the stellar mass density profile inside a fixed aperture, within $10$~kpc from the galaxy center.
} 
   {
      We first studied the stellar mass and mass-weighted density slope within $10 ~\rm{kpc}$, respectively $M_{*,10}$ and $\mathit{\Gamma_{*,10}}$, of a sample of early-type galaxies from the GAMA survey. We measured the \gmr~and its evolution over the redshift range $0.17\leq z \leq 0.37$. We then built a toy model for the merger evolution of galaxies, based on N-body simulations, to explore to what extent the observed growth in \gmr~is consistent with a dry-merger evolution scenario. 
}
   {
      From the observations, we do not detect evidence for an evolution of the \gmr relation. We put an upper limit on the redshift derivative of the normalization$~(\mu)$ and slope$~(\beta)$ of the \gmr: $|\partial \mu/\partial \log (1+z)| \leq 0.13$ and $\left|\partial \beta/\partial \log (1+z)\right| \leq 1.10$, respectively. Simulations show that most mergers induce a decrease in $\mathit{\Gamma_{*,10}}$ and an increase in $M_{*.10}$, although some show a decrease in $M_{*,10}$, particularly for the most extended galaxies and smaller merger mass ratios.
      By combining the observations with our merger toy model, we placed an upper limit on the fractional stellar mass growth of $f_M = 11.2 \%$ in the redshift range $0.17 \leq z \leq 0.37$. 
}
   {
   While our measurement is limited by systematics, the application of our approach to samples with a larger redshift baseline, particularly with a time interval $\Delta t \geq 3.2~\mathrm{Gyr}$, should enable us to detect a signal and help us better understand the late growth of ETGs. 
}
    
   \keywords{
      Galaxies: elliptical and lenticular, cD,  Galaxies: evolution, Galaxies: structure
               }

   \maketitle

\section{Introduction}\label{sec:intro}
Early-type galaxies (ETGs), which are typically elliptical in shape and evolve passively, are believed to represent the final stage of galaxy evolution. By the time they are observed, these galaxies have likely completed most of their star formation activity. However, they may still undergo further evolution via mergers and accretion. Understanding the post-quenching evolution process that ETGs experienced can provide insight into the hierarchical structure formation theory, which is fundamental in the context of the Lambda Cold Dark Matter ($\Lambda$CDM) cosmology. 
\par Observations have suggested that ETGs at higher redshift (i.e. $z\approx 2$) have significant differences from their $z \approx 0 $ counterpart, in a few aspects. For instance, ETGs at $z \approx 2$ are more compact in size \citep[e.g.][]{daddiPassivelyEvolvingEarlyType2005, toft2007, trujillo2006, trujillo2007, vandokkum2008} and have smaller color gradient, i.e. the contrast in color between the outskirt and the center is less significant than their low-redshift counterparts \citep[e.g.][]{Suess2019a,Suess2019b,Suess2020}. To be precise, the apparent size growth at fixed mass of such quiescent, compact objects is believed to be a factor of about 3 between $z = 2$ and $z = 0$ \citep[e.g.][]{fan_dramatic_2008,  van_dokkum_growth_2010,vanderwel3DHSTCANDELSEvolution2014,damjanov2019,hamadouche2022}. Combined with the fact that the star formation rate of ETGs is relatively low, this suggests that these galaxies have been undergoing a process of building up their outer envelopes since $z \approx 1.5$ via some form of merging and accretion~\citep[e.g.][]{hopkins2009,hopkins2010,van_dokkum_growth_2010}. Numerous works in the literature that are based on both observations and hydrodynamical simulations have suggested that, among various possible mechanisms, the major one should be dissipationless (dry) mergers, especially mergers with a relatively small mass ratio between the accreted galaxy and the progenitor galaxy (hereafter minor mergers). The rationale is that minor mergers can produce the largest growth in size for the same accreted mass, as well as qualitatively reproducing the evolution trend in central densities and orbital structures of these quiescent galaxies~\citep[e.g.][]{naab_minor_2009, van_dokkum_2010_hubble, oser_cosmological_2011, newman2012, hilz_how_2013, dekel_wet_2014, deugenio2023}. Recently, taking advantage of the depth and resolution of the James Webb Space Telescope Advanced Deep Extragalactic Survey \citep{Gardner_JWST,Eisenstein_JADES}, \cite{Suess2023} suggested that minor merger also have the potential to contribute to the color gradient evolution of ETGs \citep{Suess2023}.
\par Nevertheless, a quantitative investigation of the evolutionary process reveals that the minor merger scenario is insufficient for explaining the full growth of ETGs at redshift $z > 1$. Works based on HST images \citep[e.g.][]{newman2012,Belli_2015} identified satellite galaxies around massive ETGs and found that the merging event of ETGs with their surrounding satellites can only account for at most half of the entire size growth. Moreover, works based on theoretical models~\citep[e.g.][]{nipotitreu09,nipoti2009,nipoti12} also imply that it is hard to reproduce the evolution of the $R_e-M_*$ relation, i.e. the relation between the half light radius $R_e$ and the total stellar mass $M_*$, by solely invoking dry mergers. Fortunately, some possible solutions to this problem have already been proposed. Given the fact that the observed size of star-forming galaxies is on average larger than that of quiescent ones ~\citep[e.g.][]{newman2012,vanderwel3DHSTCANDELSEvolution2014,Belli_2015,KiDS_Roy}, the joining of newly quenched ex-star-forming galaxies to the population of ETGs could be a possible supplement to the size growth of ETGs ~\citep[e.g.][]{van_dokkum_1996,vandokkumMorphologicalEvolutionAges2001,carolloNEWLYQUENCHEDGALAXIES2013,fagioliMinorMergersProgenitor2016}. The existence of color gradients could also mimic the observed size growth of ETGs, thus mislead to a conclusion that the minor merger scenario is insufficient~\citep{Suess2019a,Suess2019b}. Nevertheless, detailed theoretical models that can take various possible scenarios into account and explain the entire growth of ETGs are still missing. It must also be noted that the evolution of other scaling relations (such as those involving the central stellar velocity dispersion) provide additional and complementary constraints to such models \citep{carlo2020,Nipoti2025}. 
\par Another problem arises at lower redshift, $z \leq 1$, where there is not yet consensus on the observed evolution of ETGs. The growth rate of ETGs at that time seems to be mass-dependent. \cite{vanderwel3DHSTCANDELSEvolution2014,KiDS_Roy} suggested a more rapid size growth rate for more massive ETGs ($M_* \geq 10^{11} M_\odot$), but \cite{damjanov2019} found that less massive ETGs tends to grow faster. Moreover, studies exploring the number density growth of ETGs \citep[e.g.][]{Bundy2017,Kawinwanichakij2020} imply a lack of growth in the stellar mass function in the same redshift interval, which is in general in agreement with~\cite{damjanov2019} as they both imply that massive ETGs do not grow significantly, and thus in contrast with~\cite{vanderwel3DHSTCANDELSEvolution2014} and~\cite{KiDS_Roy}. Such complexity limits our ability to draw precise conclusions on the evolution of ETGs in this redshift range. Whether the ETGs grow in this redshift range is still not determined, let alone the detailed growth mechanism. 
\par So far, the ETGs evolution has mainly been studied by means of the total stellar mass $M_*$ and the half-light radius $R_e$.
But the full light (stellar) profile contains more information than can be compressed in these two quantities.
Indeed, efforts have been made to extract the light profile to the faint outer region and study the evolution of the stellar halo~\citep[e.g.][]{tal_2011_faint,dsouza2014, Huang2018, Huang2020,spavone2021, Williams2024}. 
The outer stellar distribution is believed to preserve information about the accretion history of galaxies, with flatter light profiles being an indication of a larger fraction of accreted stellar mass.
In this work, we choose a complementary approach, by focusing on the inner region of ETGs and their evolution.

\par Following \cite{Alessandro20}, we defined the inner region as the region inside a circularized radius, $10~\rm{kpc}$, and described the stellar density profile using the mass ($M_{*,10}$) and mass-weighted density slope ($\mathit{\mathit{\Gamma_{*,10}}}$) enclosed within that aperture. The mass-weighted density slope is defined as  
\begin{equation}
    \label{eq:gammastar10}
    \mathit{\Gamma_{*,10}} = -\frac{2\pi \int_0^{10} R \frac{d\log\Sigma_*}{d\log R}\Sigma_*(R)dR}{2\pi \int_0^{10}R\Sigma_*(R)dR} = 2 - \frac{2\pi \times 10^2 \times \Sigma_*(10)}{M_{*,10}},
\end{equation} 
where radii are expressed in $\mathrm{kpc}$. These two parameters provide a good summary of the inner stellar distribution of ETGs, in the sense that by specifying their values we can predict the stellar profile inside $10 ~\rm{kpc}$ to better than $20\%$ \citep[see Fig.8 in][]{Alessandro20}. Therefore, we refer to these two parameters as the 10$~\rm{kpc}$ collar. The choice of $10~\rm{kpc}$ is arbitrary, but is a compromise between a scale that is sufficiently large to enclose a substantial fraction of the mass of a massive ETG, while not too large for this fraction to be one. A benefit of using this parameterization is that it does not suffer from extrapolation problems: at the redshifts and depths covered by our study, the surface brightness of ETGs is detected out to $10$~kpc, and therefore $M_{*,10}$ and $\Gamma_{*,10}$ can be measured directly from the data. 
On the contrary, obtaining $M_*$ and $\reff$ requires extrapolating the surface brightness model distribution out to regions that are below the sky background level. As shown by \citet{Alessandro20}, this extrapolation can be significant even at relatively low redshifts $z\sim0.2$.
\par With the help of our new robust parameterization, we measured the evolution of the inner stellar density profile of ETGs in the redshift range $0.15 \leq z \leq 0.40$. We first collected a sample of quiescent galaxies, measured their $M_{*,10}$ and $\mathit{\Gamma_{*,10}}$, and then analyzed the evolution of the \gmr. Further, to explore to what extent can the evolution of this \gmr~be explained by dry mergers, we utilized the set of simulations used in \cite{sonnenfeld2014} and originally presented in \cite{nipoti2009,nipoti12}, which contains a number of binary merger simulations with different merger mass ratios. We used $f_M$, the fractional stellar mass growth of galaxies, to represent the growth of ETGs driven by dry mergers, and built a toy model to estimate the maximum $f_M$ that can be allowed by the observed evolution in the \gmr~in the redshift range $0.15 \leq z \leq 0.40$.
\par The structure of this paper is as follows. We describe the observation sample together with the measurement of $M_{*,10}$ and $\mathit{\Gamma_{*,10}}$ in Sect.\ref{sec:observation}. In Sect.\ref{sec:3}, we present the measurement of the \gmr~using a Bayesian hierarchical method and in Sect.\ref{sec:toy}, we describe how we establish a toy model based on simulations to find a maximum fractional mass growth $f_M$ that is allowed by the observed evolution of the \gmr, in the context of dry mergers. Then we discuss the result in Sect.\ref{sec:4} and give a brief conclusion in Sect.\ref{sec:5}.
\par We assume a flat $\Lambda$CDM cosmology with $\Omega_M = 0.3$ and $H_0 = 70~\rm{km ~s^{-1}~Mpc^{-1}}$. Magnitudes are in AB units and stellar masses are in solar units.

\section{Observations}\label{sec:observation}

\subsection{Sample selection}
\par The goal of this work is to focus on the evolution of quiescent galaxies at low redshift and attempt to eliminate possible systematic effects from the extrapolation problem, by focusing on a fixed aperture size of $10~\rm{kpc}$. We want to select a sample that consists of quiescent galaxies with spectroscopic data, hence obtaining reliable measurements of redshift, stellar mass, and aperture size and distinguishing them from star-forming ones. In particular, we want the method for stellar mass measurement to be homogeneous throughout the whole sample. In addition, we also need to obtain the structural parameters of galaxies to calculate $M_{*,10}$ and $\mathit{\Gamma_{*,10}}$.
\par Therefore, we selected galaxies based on the Galaxy And Mass Assembly Data Release 4 \citep[GAMA DR4,][]{GAMA1, GAMA2,bellstedt_galaxy_2020, GAMAmain}. In practice, we used the Data Management Unit (DMU) gkvScienceCatv02 \citep{bellstedt_galaxy_2020} to select galaxies brighter than an $r$-band ProFound magnitude of $19.65$ from the GAMA DR4 Main Survey sample, which has a 95\% spectroscopic completeness. We conservatively only included galaxies in G09, G12, and G15 fields, in concern of the shallower r-band magnitude limit in the G23 field. Further, we obtained the stellar mass measurement from the DMU StellarMassesGKVv24\citep[see][]{GAMAmain}. The stellar mass estimation was obtained by applying stellar population synthesis modeling to the observed spectral energy distribution (SED) data. The fitting code is first described in \cite{Taylor2011}, but later modified so that the model-fitting was operated within a fixed wavelength range (3000 - 11000\r{A}). The fitting used the \cite{bruzual_2003} stellar evolution models with \cite{chabrier2003} stellar initial mass function (IMF), uniform metallicity, exponentially declining star formation histories and the \cite{calzetti2000} dust curve. The main reference for the measurement is \cite{Taylor2011}, while the photometry data that was used in this stellar mass estimation procedure is from \cite{bellstedt_galaxy_2020}. 
\par To select ETGs, we used the spectroscopic index $\rm{D_n}4000$ to distinguish the quiescent galaxies. In GAMA, this index is defined using the narrow band definition of \cite{Balogh99}, which is the ratio between the flux per unit frequency in $4000-4100$\r{A} and $3850-3950$\r{A}. This index is commonly used as an indicator of the age of the stellar population. According to \cite{Kauffmann2003}, the distribution of $\rm{D_n}4000$ exhibits a strong bimodality with a clear division between star-forming and quiescent galaxies at $\rm{D_n}4000 = 1.5$. Therefore, we applied a selection criterion of $\rm{D_n}4000 > 1.5$ to select quiescent galaxies. We also removed galaxies whose normalized redshift quality $nQ \leq 2$ following the suggestion by GAMA Collaboration \citep{GAMAmain}.
\par In addition, we only included galaxies in our sample that overlap with the Kilo-Degree Survey \citep[KiDS,][]{deJong2013, kuijken_fourth_2019}, to obtain surface brightness profiles of our galaxies. 
Galaxy structural parameters were measured using GalNet \citep{GaLNet2022}, which has operated a single S\'{e}rsic model fitting to the surface brightness of KiDS DR5 galaxies using its r-band photometry. After excluding galaxies with catastrophic measurements, we ended up with 66145 ETGs with measurements of spectroscopic redshift, stellar mass, and structural parameters. 

\subsection{$M_{*,10}$ and $\mathit{\Gamma_{*,10}}$ measurement}\label{sec:m10measurments}
\par To obtain $M_{*,10}$ and $\mathit{\Gamma_{*,10}}$ of the galaxies in our sample, we fit the image of galaxies with a Point Spread Function (PSF)-convolved S\'{e}rsic model, then derived the $M_{*,10}$ and $\Gamma_{*,10}$ based on the best-fitting model.
While the total mass $M_*$ and half-light radius $R_e$ derived by fitting a S\'{e}rsic model suffers from the extrapolation problem discussed in \Sref{sec:intro}, the quantities $M_{*,10}$ and $\mathit{\Gamma_{*,10}}$ do not. This is because, as we show below, the surface brightness profile in the inner $10$~kpc is well constrained by our data, and is well described by our best-fitting S\'{e}rsic model.
Although, in principle, $M_{*,10}$ and $\Gamma_{*,10}$ could be determined in a non-parametric way, by directly measuring the surface brightness profile, our model-based approach allows us to more conveniently deblend the galaxies of interest from contaminants and to deconvolve the surface brightness profile by the PSF \citep[see][for details]{GaLNet2022}.

The S\'{e}rsic profile \citep{Sersic1968} can be written as:
\begin{equation}
    I(R) = I_0 \exp\left\{-b_n\left(\frac{R}{R_e}\right)^{1/n}\right\},
    \label{SB}
\end{equation}
with
\begin{equation}
    R^2 = qx^2 + \frac{y^2}{q}.
\end{equation}
Here, $q$ is the axis ratio, $n$ is the S\'{e}rsic index while $R$ is the circularized radius where $x$ and $y$ are Cartesian coordinates with origin at the center of galaxies. We use the symbol $x$ to denote the axis that is aligned with the semi-major axis of the ellipse while using $y$ to denote the axis aligned with the semi-minor axis. The effective radius $\reff$ is circularized as well. The light enclosed within a circularized aperture of radius $R$ is given by: 
\begin{equation}
    \label{eq:light}
    L(<R) = 2\pi n\cdot I_0R_e^2 \cdot \frac{1}{\left(b_n\right)^{2n}}\cdot \gamma\left[2n,b_n \left(\frac{R}{R_e}\right)^{\frac{1}{n}}\right].
\end{equation}
The total light can be obtained by letting $R \rightarrow +\infty$,
\begin{equation}
    L_{tot} = 2\pi n\cdot I_0R_e^2 \cdot \frac{1}{\left(b_n\right)^{2n}}\cdot \Gamma\left(2n\right).
\end{equation}
Here $\Gamma$ is the gamma function, $\gamma$ is the lower incomplete gamma function and $b_n$ is a constant that ensures the light enclosed within the effective radius $R_e$ is half of the total light:
\begin{equation}
    L_{tot} = 2 L(<R_e).
\end{equation}
Therefore, $b_n$ can be calculated by solving
\begin{equation}
    \Gamma(2n) = 2\gamma(2n, b_n).
\end{equation}
\par
\begin{figure*}
    \centering
    \includegraphics[width=0.8\linewidth]{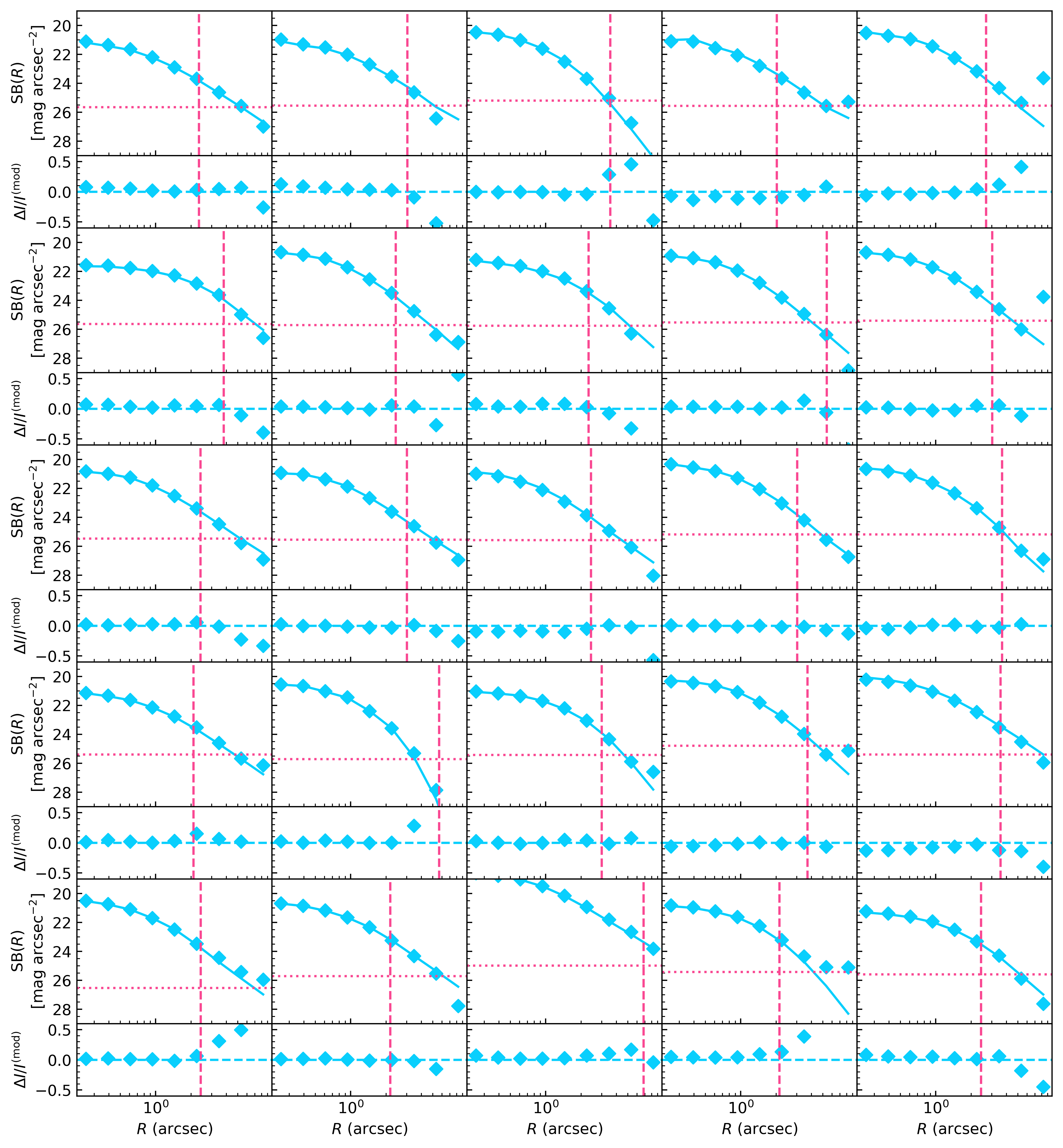}
    \caption{ The surface brightness profile of 25 galaxies. We selected these galaxies after visually inspecting 69 galaxies in one KiDS tile and randomly choosing 25 from the 62 that do not have bright companions. The upper panel shows the surface brightness as a function of the circularized radius. The blue solid line shows the best-fitting S\'{e}rsic model convolved with PSF, calculated using GalNet structural parameters. The blue diamonds are directly measured from the KiDS r-band image. The lower panel shows the difference between the two measurements. The vertical dashed pink line shows the corresponding angular size of 10 $\mathrm{kpc}$ of each galaxy, the horizontal line shows the noise level of the sky.} 
    \label{fig:surface_brightness}
\end{figure*}

With the purpose of demonstrating the accuracy of our S\'{e}rsic fits in the inner $10\rm{kpc}$ region, we show in Fig.\ref{fig:surface_brightness} a comparison between the PSF-convolved best-fitting model surface brightness profiles and the circularized surface brightness profiles measured in radial bins directly from the data, for 25 galaxies. We selected these galaxies randomly, while discarding objects with bright companions. We only applied this additional isolation cut for the purpose of the illustration of \Fref{fig:surface_brightness}. As shown by \cite{GaLNet2022}, our fitting method can deal with objects with overlapping surface brightness distributions, and such objects are included in our analysis. On the other hand, the presence of neighbors complicates the procedure of measuring the surface brightness profile directly from the data, and with it our comparison. 
The observed light profile can be reproduced by the best-fitting S\'{e}rsic model inside $10~\mathrm{kpc}$ (denoted by the vertical dashed pink line) with a typical accuracy of $10\%$. Residuals are thus much smaller than the uncertainties on the stellar mass-to-light ratio, which are of the order of 30\%. Therefore our measurements of $M_{*,10}$ and $\Gamma_{*,10}$ are largely immune to the impact of systematics related to the determination of galaxy structural parameters. Additionally, we measured the median fractional error across the 25 galaxies, finding a value of $1.2\%$ for the surface brightness at $10~\mathrm{kpc}$ and $1.5\%$ for the enclosed light within the same radius. These results confirm that the best-fitting model provides an unbiased reconstruction of the surface brightness profile, thereby ensuring the reliability of the derived parameters $M_{,10}$ and $\Gamma_{,10}$.
\Fref{fig:surface_brightness} also shows that the radius at which the surface brightness of our galaxies falls below the sky background rms level (horizontal dashed line) is typically larger than $10$~kpc.
This means that $M_{*,10}$ and $\Gamma_{*,10}$ are indeed determined directly from the data and thus do not suffer from the extrapolation problem.

\par In our work, we assume that there is no stellar mass-to-light ratio gradient inside one galaxy, hence the stellar mass profile can be easily obtained via the light profile. We discuss the possible implications of this choice in Sect.~\ref{sec:systematic}. 
We have obtained the stellar mass estimate for each galaxy from GAMA together with the light used in the SPS model fitting process \citep{GAMAmain}, which enables us to calculate the mass-to-light ratio $\Upsilon_*$. The mass profile can be easily obtained by multiplying Eq.~\ref{SB} and Eq.~\ref{eq:light} by the stellar mass-to-light ratio $\Upsilon_*$: 
\begin{equation}
    \label{eq:sigma}
    \Sigma_*(R) = \Upsilon_* I_0 \exp\left[-b_n\left(\frac{R}{R_e}\right)^{1/n}\right], 
\end{equation}
\begin{equation}
    \label{eq:mass}
    M_*(<R) = \Upsilon_* 2\pi n\cdot I_0R_e^2 \cdot \frac{1}{\left(b_n\right)^{2n}}\cdot \gamma\left[2n,b_n \left(\frac{R}{R_e}\right)^{\frac{1}{n}}\right].
\end{equation}
By simply substituting $R = 10~\rm{kpc}$ into Eq.~\ref{eq:mass}, we obtained $M_{*,10}$, while we obtained $\mathit{\Gamma_{*,10}}$ by combining Eq.~\ref{eq:sigma} with Eq.~\ref{eq:gammastar10}. 
\par The stellar mass-to-light ratio $\Upsilon_*$ and the S\'{e}rsic structural parameters $R_e$ and $n$ are measured by two different surveys: $\Upsilon_*$ ratio from GAMA, and $R_e$ and $n$ from KiDS.\@ The total flux of a galaxy used in KiDS S\'{e}rsic modeling might be different from that used in GAMA stellar mass estimate, so we need to take this difference into account when obtaining $M_{*,10}$ and $\mathit{\Gamma_{*,10}}$. Nevertheless, as we have assumed that there is no gradient in the stellar mass-to-light ratio, the measured value of $\Upsilon_*$ does not depend on the modeling choice. Therefore, although we can only obtain $\Upsilon_*$ from the GAMA survey, we believe it is reasonable to use it to calculate $M_{*,10}$ and $\mathit{\Gamma_{*,10}}$ together with the structure parameter measured by KiDS.
\subsection{Completeness}\label{sec:completeness}
In order to minimize potential biases related to selection effects, we built a volume-limited sample in $M_{*,10}$. The GAMA DR4 Main Survey sample, from which the quiescent galaxies were selected, is flux limited, with a 95\% completeness limit down to r-band ProFound magnitude of 19.65, which correspond to a critical flux $F_{crit} \approx 5.01 \times 10^{-5} \text{Jy} $. To obtain a $M_{*,10}$ complete sample, we needed to translate this completeness limit in flux to a limit in $M_{*,10}$.
\par We used the $M_{*,10}$-to-flux ratio to operate this translation. At a given redshift, the ratio between $M_{*,10}$ and the total flux $F$ is not a fixed value, but spread over a relatively wide range. We made narrow redshift bins, measured the distribution of $M_{*,10} / F$ in each bin and found the critical value $M_{*,10} / F|_{crit}$, where the cumulative probability reaches $95\%$. Multiplying $M_{*,10} / F|_{crit}$ by the critical flux $F_{crit}$, we obtained the $M_{*,10}$ limit at that redshift bin. Here we made an implicit assumption that the ratio $M_{*,10} / F$ depends neither on $M_{*,10} $ nor on $F$.
\begin{figure}
    \centering
    \includegraphics[width=\linewidth]{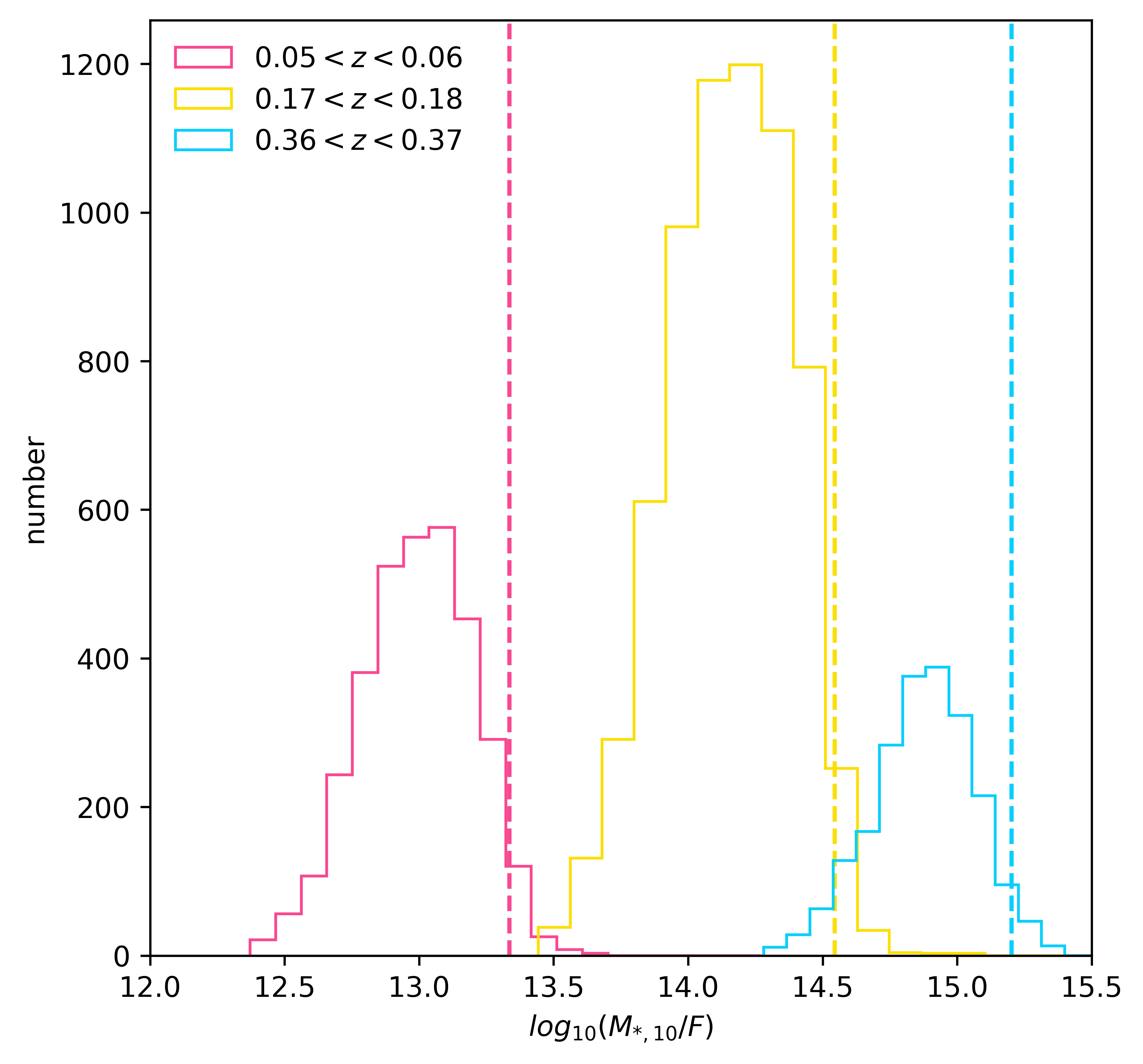}
    \caption{The distribution of $M_{*,10} / F$ in three different narrow redshift bins. The vertical dashed line marks the $95\%$-ile distribution in each bin. By multiplying this ratio by the flux corresponding to the r-band magnitude limit $r_{crit} = 19.65$, we then obtain the $M_{*,10}$ limit at that redshift bin.}
    \label{fig:m2f}
\end{figure}
\par Fig.~\ref{fig:m2f} illustrates the procedure, which shows the distribution of $M_{*,10} / F$ in three different redshift bins, with dashed lines indicating the $95\%$-ile of each bin, i.e. the critical $M_{*,10}$-to-flux ratio $M_{*,10}/F|_{crit}$. The $M^{crit}_{*,10} $ in these three bins can thus be calculated by $M^{crit}_{*,10} = M_{*,10}/F|_{crit} \times F_{crit} $ using the value of $M_{*,10}/F|_{crit}$ in each bin. Applying this procedure iteratively in each redshift bin results in the full $95\%$ completeness limit in $M_{*,10}$ as a function of redshift $z$. We used a logarithmic formula to fit this limit as a function of redshift
\begin{equation}
    f(z) = A\log z + B.
\end{equation}
The best-fitting parameters are $A = 0.87 \pm 0.0001$ and $B = 11.74 \pm 0.0003$.

\par
\begin{figure}
    \centering
    \includegraphics[width = \linewidth]{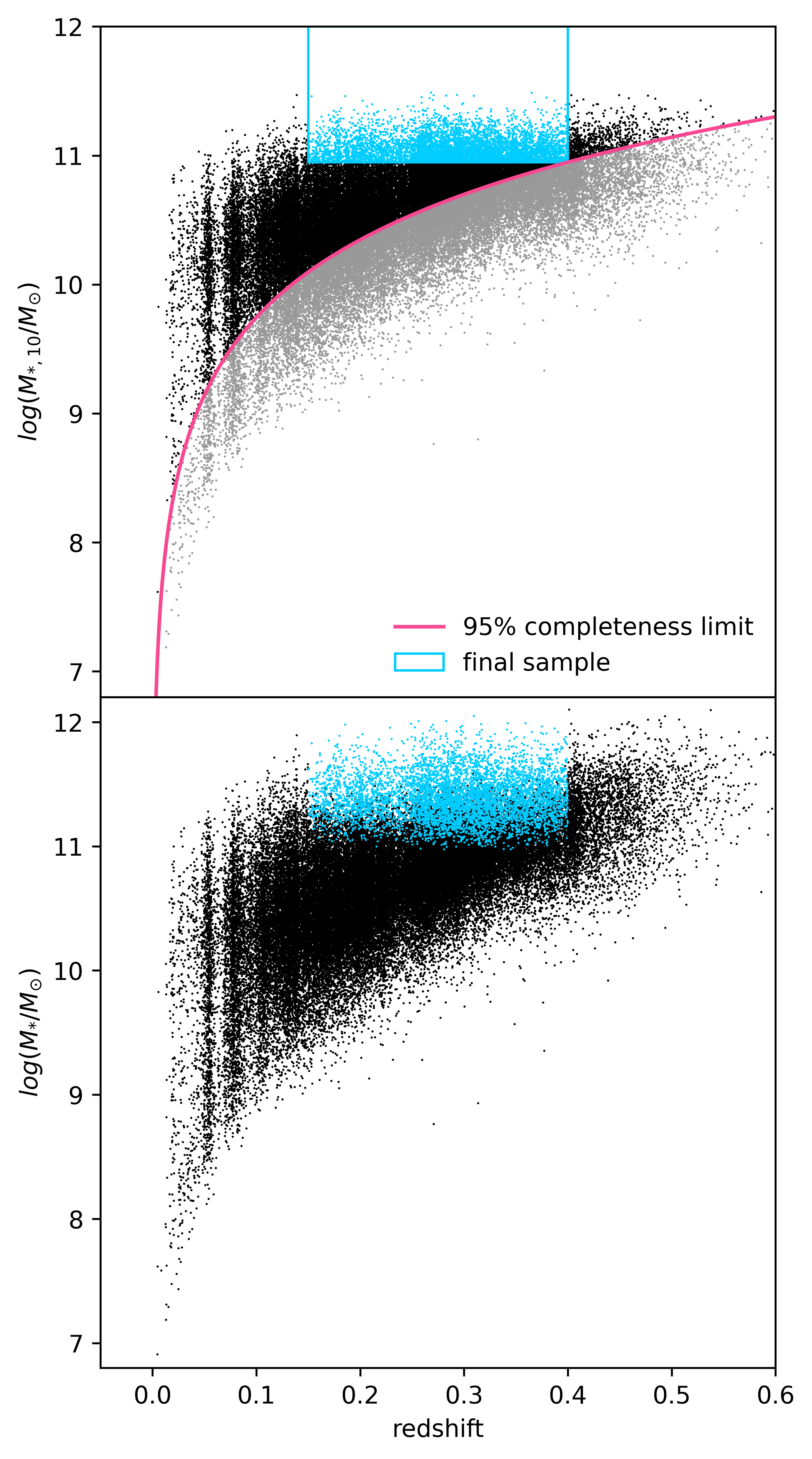}
    \caption{The distribution of our galaxy sample in $z- M_{*,10}$ space (top) and $z - M_*$ space (bottom). The $95 \% $ completeness $M_{*,10} $ limit as a function of redshift of our ETG sample is shown in pink solid line. Galaxies above this pink line are adopted in our complete sample, shown in black dots, and the Gray dots that lies below are those being excluded. We further define a lower and upper redshift limit of $z \geq 0.15$ and $z \leq 0.40$ respectively, to build the final sample and shown in blue dots.}
    \label{fig:completeness_cut}
\end{figure}

 The upper panel of Fig.~\ref{fig:completeness_cut} shows the distribution of our sample in $M_{*,10} - z$ space, with the pink solid line showing the $95\%$ completeness limit. Galaxies whose $M_{*,10}$ is larger than the limit have a $95\%$ probability to be included in our sample. We thus excluded those galaxies whose $M_{*,10}$ is lower than this limit in our sample and show these galaxies as gray dots. Galaxies shown as black dots are included to build our complete sample. The lower panel shows all our sample in $M_* - z$ space with black dots.

\par We only focused on massive quiescent galaxies whose $M_{*,10} \geq 10^{10.9} M_{\odot}$, thus a corresponding upper limit in redshift $z = 0.4$ is placed. We further chose to set a lower limit on the redshift range $z \geq 0.15$, in order to increase the S/N ratio with a reliable large sample size. The final sample that we used to investigate the evolution contains 5690 ETGs, and is shown by cyan dots in the upper panel of Fig.~\ref{fig:completeness_cut}. Also, we show the distribution of our final sample in $z - M_*$ space in the lower panel in cyan dots, with the others shown in black. The $5$th percentile of the stellar mass distribution of our final sample is $\log M_* = 11.1$. The distribution of our sample in $M_{*,10},~\mathit{\Gamma_{*,10}}$ and $z$ is shown in Fig.~\ref{fig:data_feature}.
\section{\gmr}\label{sec:3}
\subsection{Model}
\par Having obtained the measurements of $M_{*,10}$ and $\mathit{\Gamma_{*,10}}$, we then measured the \gmr. We believe that this relation can provide us with valuable information on the growth of ETGs, complementary to the traditional $R_e - M_*$ scaling relation. In order to infer the distribution of $\mathit{\Gamma_{*,10}}$ as a function of $M_{*,10}$, we fitted a model with a Bayesian hierarchical method. Each galaxy included in our sample can be described by a set of quantities $\{ \log M_{*,10}, \mathit{\Gamma_{*,10}}, z \}$. The true value of these quantities, hereafter represented by $\mathbf{\Theta}$, are drawn from a probability distribution that correlates with the observed data $\mathbf{d} = \{ \log M_{*,10}^{obs}, \mathit{\Gamma_{*,10}}^{obs}, z^{obs} \}$, and is in turn described by a set of hyper-parameters $\mathbf{\Phi}$. We want to obtain the posterior distribution of the hyper-parameters $\mathbf{\Phi}$ given the observed data $\mathbf{d}$: 
\begin{figure}
    \centering
    \includegraphics[width=\linewidth]{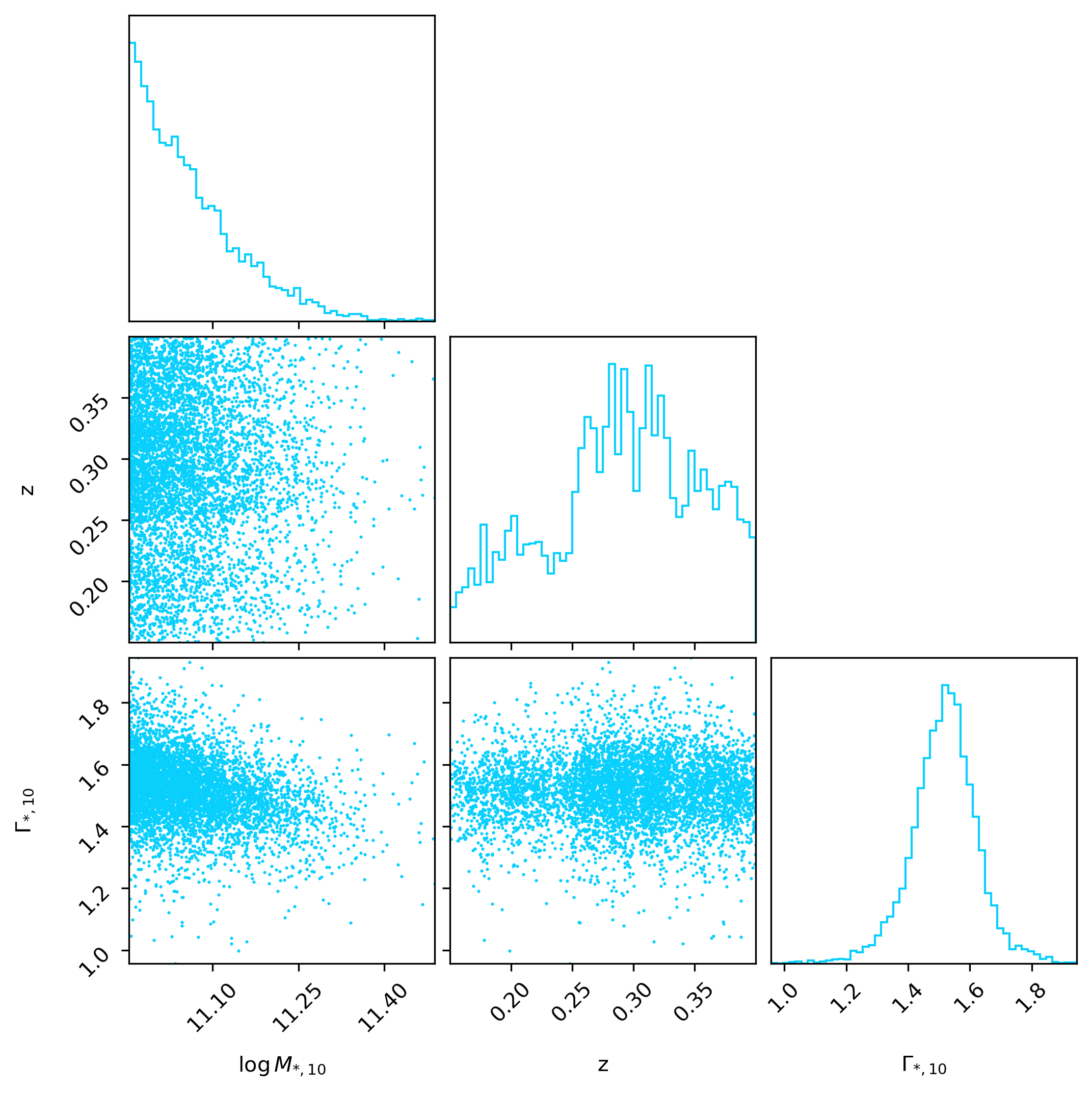}
    \caption{The distribution of our 5690 ETGs in $M_{*,10}$, $\mathit{\Gamma_{*,10}}$ and redshift $z$.}
    \label{fig:data_feature}
\end{figure}
\begin{equation}
    \mathrm{P}(\mathbf{\Phi}|\mathbf{d}) \propto  \mathrm{P}(\mathbf{\Phi}) \mathrm{P}(\mathbf{d}|\mathbf{\Phi}),
\end{equation}
thus we marginalize over every possible value of $\mathbf{\Theta}$ in the likelihood part: 
\begin{equation}
    \label{Eq:likelihood}
    \mathrm{P}(\mathbf{d}|\mathbf{\Phi}) = \int \mathrm{P}(\mathbf{d}|\mathbf{\Theta})\mathrm{P}(\mathbf{\Theta}|\mathbf{\Phi})d\mathbf{\Theta}.
\end{equation}
\par The likelihood term on the right-hand side of Eq.~\ref{Eq:likelihood} can be rewritten as
\begin{equation}\label{eq:pdphi}
    \mathrm{P}(\mathbf{d}|\mathbf{\Theta}) = \mathrm{P}(\log M^{obs}_{*,10}|\log M_{*,10})\mathrm{P}(\Gamma^{obs}_{*,10}|\mathit{\Gamma_{*,10}})\mathrm{P}(z^{obs}|z).
\end{equation}

\par $\mathrm{P}(\mathbf{d}|\mathbf{\Theta})$ corresponds to the uncertainties of $\log M_{*,10}$, $\mathit{\Gamma_{*,10}}$ and $z$. In this work, we only consider the uncertainty from the measurement of the stellar mass. Uncertainties from other sources are sufficiently small to be ignored. For instance, $\mathrm{P}(z^{obs}|z)$ is one negligible term as it is only affected by the uncertainties on the spectroscopic redshift. The term $\mathrm{P}(\Gamma^{obs}_{*,10}|\mathit{\Gamma_{*,10}})$ can also be ignored: $\mathit{\Gamma_{*,10}}$ does not depend on stellar mass, the measurement of the surface brightness profile is the only source of uncertainty, which is negligible as shown in Fig.~\ref{fig:surface_brightness}. The remaining term of $\mathrm{P}(\mathbf{d}|\mathbf{\Theta})$ is $\mathrm{P}(\log M^{obs}_{*,10}|\log M_{*,10})$, which we assume to be a truncated Gaussian distribution, as we only selected galaxies whose $\log M_{*,10}^{obs} > \log M_{*,10,\min}$. When $\log M_{*,10}^{obs} > \log M_{*,10,\min}$, we have:
\begin{equation}
    \begin{aligned}
        \mathrm{P}\left(\log M_{*,10}^{obs}|\log M_{*,10}\right) &= \frac{\mathcal{A}[\log M_{*,10}]}{\sqrt{2\pi\sigma^2_{M_{*,10,obs}}}} \\
        &\times \exp\left\{-\frac{[\log M_{*,10} - \log M_{*,10}^{obs}]^2}{2\sigma_{M_{*,10, obs}}^2}\right\}, 
    \end{aligned}       
\end{equation}
otherwise 
\begin{equation}
    \mathrm{P}\left(\log M_{*,10}^{obs}|\log M_{*,10}\right) = 0.
\end{equation}
The factor $\mathcal{A}[\log M_{*,10}]$ is a normalization constant that ensures that the probability of obtaining any $\log M_{*,10}$ measurement larger than $\log{M_{*,10,\mathrm{\min}}}$, given that a galaxy is part of our sample, is one. Therefore, the following equation for $\mathcal{A}$ needs to be satisfied:
\begin{equation}
    \begin{aligned}
        \int_{\log M_{*,10,\min}}^\infty d\log M_{*,10}^{obs}&\frac{\mathcal{A}[\log M_{*,10}]}{\sqrt{2\pi\sigma^2_{M_{*,10,obs}}}} \\
        &\times \exp\left\{-\frac{[\log M_{*,10} - \log M_{*,10}^{obs}]^2}{2\sigma_{M_{*,10, obs}}^2}\right\} = 1.
    \end{aligned}
\end{equation}
\par The second part on the right-hand side of Eq.~\ref{Eq:likelihood} can be rewritten as
\begin{equation}
    \label{eq:second}
    \mathrm{P}(\mathbf{\Theta}|\mathbf{\Phi}) =  \mathrm{P}(\log M_{*,10},z|\mathbf{\Phi})\mathrm{P}(\mathit{\Gamma_{*,10}}|\mathbf{\Phi},\log M_{*,10},z).
\end{equation}
This part is the core of our model, as it not only contains the distribution of $\log M_{*,10}$ but also the relation between $\log M_{*,10}$ and $\mathit{\Gamma_{*,10}}$. We want to fully capture both the average \gmr~and its redshift evolution. We chose a model-independent way to describe the redshift evolution, aiming at avoiding the potential bias that might be introduced by an inadequate choice of the functional form. We separated the sample into different redshift bins and, for each bin, fit the model while ignoring the redshift dependence of $\mathit{\Gamma_{*,10}}$.
\par The description of this model is as follows. Without the redshift dependence, Eq.~\ref{eq:second} can be simplified as
\begin{equation}
    \label{eq:sec_noz}
    \mathrm{P}(\mathbf{\Theta}|\mathbf{\Phi}) =  \mathrm{P}(\log M_{*,10}|\mathbf{\Phi})\mathrm{P}(\mathit{\Gamma_{*,10}}|\mathbf{\Phi},\log M_{*,10}).
\end{equation}
\par  For the distribution of $\log M_{*,10}$, we use a skewed normal distribution, inspired by the functional form used by~\cite{carlo2020}: 
\begin{equation}
    \label{eq:logM_distribution}
    \begin{aligned}
        \mathrm{P}\left(\log M_{*,10}|\Phi\right) &= \frac{1}{\sqrt{2\pi\sigma^2_M}}\exp\left\{-\frac{(\log M_{*,10} - \mu_M)^2}{2\sigma_M^2}\right\} \\
        &\times \mathcal{E}\left(\log M_{*,10} | \Phi\right),
    \end{aligned}
\end{equation} 
where
\begin{equation}
    \mathcal{E}\left(\log M_{*,10} | \Phi\right) = 1 + \erf \left(\alpha_M\frac{\log M_{*,10} - \mu_M}{\sqrt{2\sigma_M}}\right).
\end{equation}
$\mathrm{P}\left(\log M_{*,10}|\Phi\right)$ serves as a prior on $\log M_{*,10}$. Although a functional form of skewed Gaussian distribution might not be a perfect description of the $\log M_{*,10}$ distribution, this issue does not significantly bias the posterior distribution. Adopting a Schechter function for the prior can only shift the mean and the standard deviation of the posterior distribution no more than $0.01$ dex and $1\%$, respectively. The uncertainty on $\log M_{*,10}$ is smaller than the width of the $\log M_{*,10}$ distribution, hence the posterior distribution is dominated by the second term, i.e. the likelihood part. The influence of the prior is therefore minimal. 
In addition, 
our model automatically corrected the so-called Eddington bias. 
\par For the distribution of $\mathit{\Gamma_{*,10}}$, we used a Gaussian model:
\begin{equation}\label{eq:gamma_distribution}
    \mathrm{P}\left(\mathit{\Gamma_{*,10}}|\Phi,\log M_{*,10}\right) = \frac{1}{\sqrt{2\pi\sigma^2_{\Gamma}}}\exp\left\{-\frac{(\mathit{\Gamma_{*,10}} - \mu_{\Gamma})^2}{2\sigma_{\Gamma}^2}\right\}.
\end{equation}
We descried the relation between $\log M_{*,10}$ and $\mathit{\Gamma_{*,10}}$ by a linear dependence of $\mu_\Gamma$ on $\log M_{*,10}$:
\begin{equation}
    \label{eq:mg_relation}
    \mu_{\Gamma} = \mu_{\Gamma,0} + \beta\left(\log M_{*,10} - \log M^{piv}_{*,10}\right).
\end{equation}
Here we use a pivot value $\log M_{*,10} = 11.04 $, which is the median value of $M_{*,10}$ of our sample.
\par Therefore, in this model, the full distribution of $\Theta$ given $\mathbf{\Phi}$ is governed by the following set of hyperparameters:
\begin{equation}
    \mathbf{\Phi} = \{ \mu_M, \sigma_M, \alpha_M, \mu_{\Gamma,0}, \beta, \sigma_{\Gamma} \}.
\end{equation}

\subsection{Fitting result}\label{sec:conserve}

\begin{figure}
    \centering
    \includegraphics[width=\linewidth]{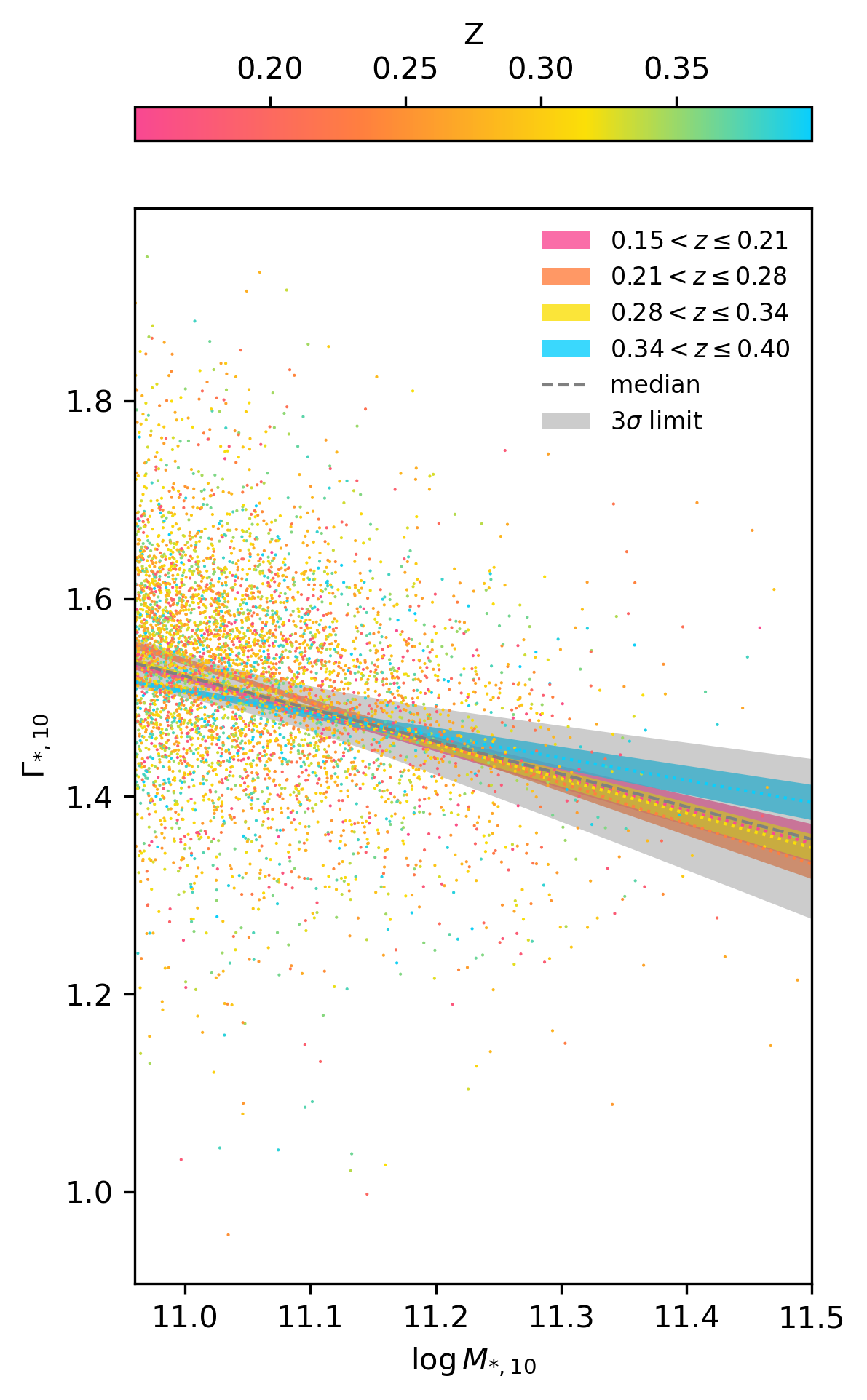}
    \caption{The \gmr~of our observation sample fitted using a Bayesian hierarchical model. For each redshift bin, the median value of this relation is shown in the dotted line, while the shaded region shows the $68\%$ credible interval. 
The gray region indicates our estimated $3-\sigma$ systematic uncertainty.
}
\label{fig:mg_relation}
\end{figure}

\begin{figure*}
\sidecaption
    \includegraphics[width = 12cm]{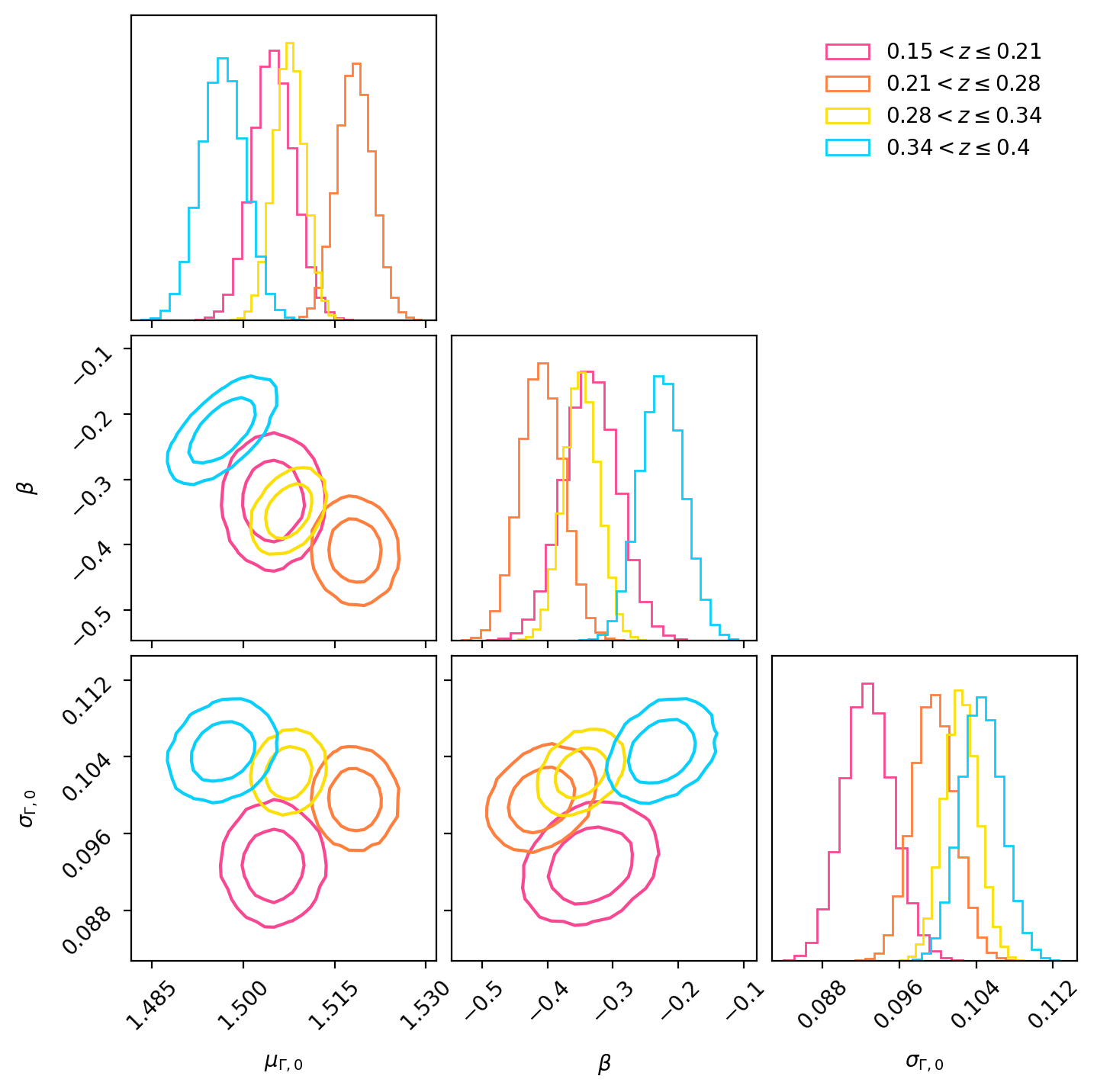}
    \caption{Posterior distribution of the hyperparameters in the Bayesian hierarchical model which correspond to the \gmr. The contours represent the $68\%$ and $95\%$ credible interval.}
    \label{fig:bin_by_bin_corner}
\end{figure*}

We binned our sample into four redshift bins of equal width and then sampled the posterior distribution of the hyperparameters in each bin. Table~\ref{tab:mg_relation} shows the median value of the marginal posterior probability in each parameter. To illustrate the \gmr, which is described by Eq.~\ref{eq:mg_relation}, we plot $\mu_\Gamma$ as a function of $\log M_{*,10}$ in Fig.~\ref{fig:mg_relation} as a dotted line, while the shaded region represents the $68\%$ credible interval, according to the posterior probability. We further show the posterior distribution of these hyperparameters that govern the \gmr~in Fig.~\ref{fig:bin_by_bin_corner}, with four different colors representing the four redshift bins. 
\begin{table}
    \renewcommand\arraystretch{1.5}
    \centering
    \caption{Median values of the marginal posterior probability of the model parameters in each redshift bin.}
    \label{tab:mg_relation}
    \resizebox{\columnwidth}{!}{
    \begin{tabular}{ccccccc}
        \hline
        redshift & $\mu_M$ & $\sigma_M$ & $\alpha_M$ & $\mu_{\Gamma,0}$ & $\beta$ & $\sigma_{\Gamma}$ \\
        \hline
        0.15-0.21 & 10.92 & 0.16 & 6.34 & $1.50$ & $-0.33$ & 0.10 \\
        0.21-0.28 & 10.94 & 0.15 & 4.34 & $1.52$ & $-0.40$ & 0.10 \\
        0.28-0.34 & 10.93 & 0.14 & 2.64 & $1.51$ & $-0.35$ & 0.10 \\
        0.34-0.40 & 10.89 & 0.15 & 2.09 & $1.50$ & $-0.23$ & 0.11 \\
        \hline
    \end{tabular}}
\end{table}

\par Fig.~\ref{fig:mg_relation} exhibits an anti-correlation between $\mathit{\Gamma_{*,10}}$ and $M_{*,10}$. This is qualitatively in agreement with the traditional $R_e - M_*$ scaling relation: galaxies with larger stellar mass tend to have larger size, i.e. larger $R_e$. For simplicity, assuming that all the galaxies have the same S\'{e}rsic index $n$, the S\'{e}rsic profile becomes steeper going outwards. For a less massive galaxy, 10 $\mathrm{kpc}$ includes more the steeper parts of the profile than for a more massive galaxy. Qualitatively this should remain true also with a spread in S\'{e}rsic index.
The relation in the highest redshift bin shows a significant difference compared with the other three bins, both in the slope and in the normalization. The other three redshift bins are in agreement at the $1 \sigma$ level. However, the evolution trend in these four bins is non-monotonic, both in the slope (reflected by $\beta$) and in the normalization (reflected by $\mu_{\Gamma,0}$). 
\par As we think that the post-quenching evolution should mainly be driven by dry mergers, we believe that the evolution of the \gmr~should be monotonic. Given this strange observed feature and our prior belief, it is difficult to interpret these observations. We believe that there are some systematic effects in the data that we have not taken into account: for instance, redshift-dependent inaccuracies in the stellar population synthesis models, the existence of a color gradient or the progenitor bias. We further discuss these possible systematic effects in Sect.~\ref{sec:systematic}. 

\par To estimate the amplitude of our systematic uncertainties, we assumed that the intrinsic \gmr~does not evolve, and attribute the differences among the four \gmr s to measurement errors. We calculated the median \gmr~of the four bins, the normalization and the slope of which are $\overline{\mu_{\Gamma,0}} = 1.51$ and $\overline{\beta} = -0.33$, respectively. We considered this median \gmr~as the underlying intrinsic \gmr. Then, we took the standard deviation of the normalization and the slope of the \gmr~in the four bins, which are $\sigma_\mu = 0.007$ and $\sigma_\beta = 0.05$, as estimates of the $1-\sigma$ amplitude of the uncertainty. 
With this, we considered a linear model for the redshift evolution of the \gmr,

\begin{equation}\label{eq:z_mu}
    \mu_{\Gamma} = \mu_{\Gamma,0} + \beta\log M_{*,10} + \zeta_\mu\log(1+z),
\end{equation}
with a redshift-dependent slope
\begin{equation}\label{eq:z_beta}
    \beta = \beta_0 + \zeta_\beta\log(1+z),
\end{equation}
\par The two parameters $\zeta_\mu$ and $\zeta_\beta$, which are the redshift derivatives of $\mu_{\Gamma,0}$ and $\beta$, are thus constrained to be $\left|\zeta_\mu\right| \leq 0.13$ and $\left|\zeta_\beta\right | \leq 1.10$ at the $3-\sigma$ level. We will further discuss the implications of this limit in Sect.~\ref{sec:toy_model}.


\section{Toy model}\label{sec:toy}
In Sect.~\ref{sec:3}, we measured the \gmr~as well as its redshift evolution. We want to know, assuming that all the evolution of ETGs is induced by dry mergers, to what extent can this scenario be consistent with our observation. We used the fractional stellar mass growth $f_M$ to represent the evolution of ETGs, and then built a toy model to find a maximum $f_M$ that is allowed by the observed evolution of the \gmr, in the context of dry mergers. 

\subsection{Simulations}\label{sec:simulation}

In order to establish a toy model, the first thing to do is to understand the impact of mergers on ETGs, particularly on $M_{*,10}$ and $\mathit{\Gamma_{*,10}}$. In this work, we chose to look at N-body simulations to get the answer. We utilized a collection of 8 sets of dissipationless binary-merger simulations which have been used in \cite{sonnenfeld2014}~\citep[see also][]{nipoti2009,nipoti12} to construct a dry-merger model. 
\par A typical simulation dataset consists of two progenitor galaxies and one remnant. The two progenitors, i.e. main galaxy and satellite galaxy, in each set of simulations are composed of dark matter and stellar mass, both in a spherically symmetric distribution. The stellar profile is described by a $\gamma$ model \citep{Dehnen93,Tremaine94} with $\gamma = 1.5$: 
\begin{equation}
    \label{eq:profile_stellar}
    \rho_*(r) = \frac{3-\gamma}{4\pi}\frac{M_*r_*}{r^\gamma (r+r_*)^{4-\gamma}},
\end{equation}  
where $M_*$ is the total stellar mass and $r_*$ is a scale radius.
\par The dark matter halo is described by a truncated NFW profile \citep{NFW},
\begin{equation}
    \label{eq:profile_halo}
    \rho_{DM}(r) = \frac{M_{DM,0}}{r(r+r_s)^2} \exp\left[-\left(\frac{r}{r_{vir}}\right)^2\right],
\end{equation}
where $r_s$ is the scale radius, $M_{DM,0}$ is a reference mass and $r_{vir} = cr_s$ is the virial radius ($c$ is the concentration). As the sum of the two components, the total mass density profile $\rho(r) = \rho_*(r) + \rho_{DM}(r)$ is nearly isothermal. The $\gamma'$, which is the logarithmic slope of the total mass profile over the radial range $r < 0.5R_{eff}$, in the considered sets of simulations lies in the range of $1.97 - 2.03$, in order to match the strong lensing measurement by \cite{auger2010}. The remnant galaxy is the collection of dark matter particles and stellar particles that are bound, after the stellar component of the merging system is fully virialized.
\par The 8 sets of simulations have 3 different merger mass ratios $\xi$, thus they can be used to probe both minor mergers and major mergers. In particular, 3 sets of simulations with 6 runs have $\xi = 0.2$, 3 sets have $\xi = 0.5$ while the other 2 sets have $\xi = 1$. The other differences among the 8 sets of simulations occur in the concentration $c$, the scale ratio $r_{s}/R_{eff}$, and the stellar-to-dark matter mass ratio $M_*/M_h$ of the progenitor galaxies. The details of these differences are reported in Table 1 of \cite{sonnenfeld2014}. Each set contains two runs of simulations with nearly identical parameter settings except for their different orbital angular momentum, in order to take both head-on and off-axis encounters into consideration. All orbits are parabolic. 
\par A limitation of choosing these three $\xi$ values is that mergers with a small merger mass ratio ($\xi < 0.1$) are not considered. Taking advantage of the deep image of JWST, \cite{Suess2023} suggested that the small mass ratio merger can also contribute to the growth of ETGs. However, recent work of \cite{Nipoti2025} indicates that the effect of these `mini mergers' may be different from mergers with the mass ratios we have considered in this work. Future efforts should be made to extend our knowledge about mergers down to lower mass ratios.
\par
\begin{figure*}
    \centering
    \includegraphics[width=\linewidth]{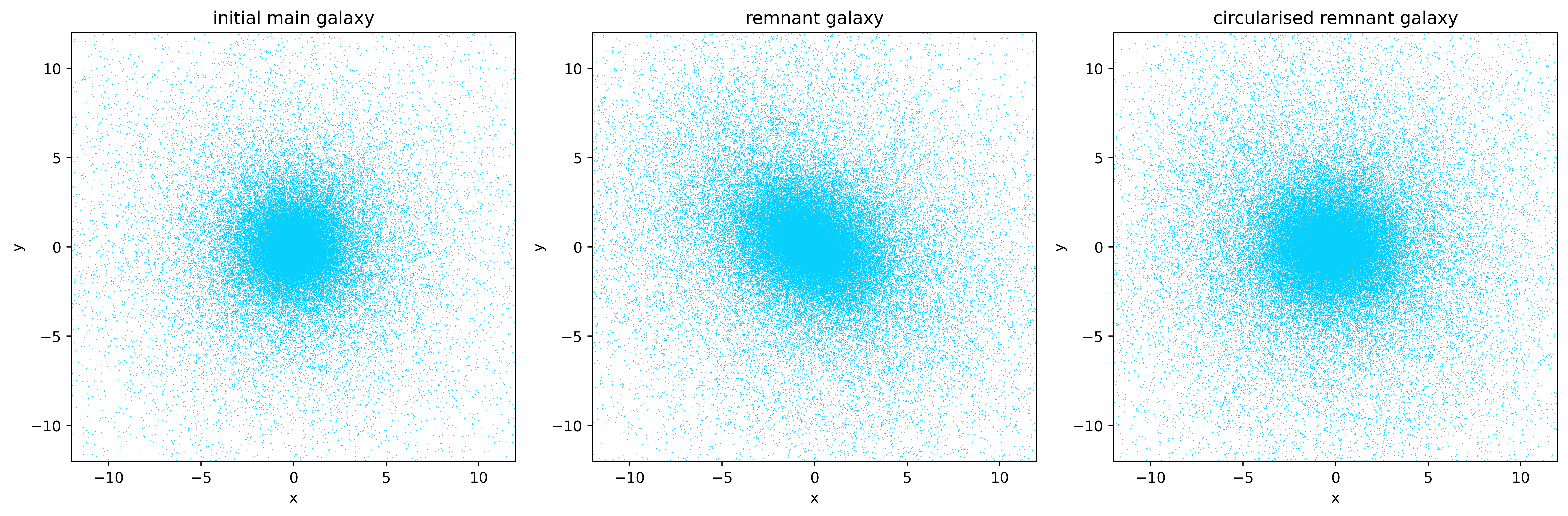}
    \caption{\label{fig:mock_image} Projected stellar particles of the main and remnant galaxies from a binary merger simulation. The left panel shows the initial main galaxy, while the middle and right panels are the uncircularized and circularized remnant galaxy respectively. Length are in arbitrary units.}
\end{figure*}

 To investigate the effect on the two parameters $M_{*,10}$ and $\mathit{\Gamma_{*,10}}$, we focused on the main progenitor galaxy and the remnant galaxy. We projected the initial and remnant galaxies onto the $x-y$ plane, which is parallel to the merger orbital plane, so as to generate mock observational images. The mock images are shown in Fig.~\ref{fig:mock_image} in the left and middle panels. The initial main galaxy is spherical, thus its image is a circle. The remnant galaxy does not have this symmetry, hence we can observe that its image is an ellipse. In order to make a fair comparison, we then circularized the remnant ellipse. In particular, we measured the quadrupole moment of the remnant galaxy image and then stretched and rotated the image according to this quadrupole moment until the image became a circle, as shown in the right panel of Fig.~\ref{fig:mock_image}.
\begin{figure*}
    \centering
    \includegraphics[width=\linewidth]{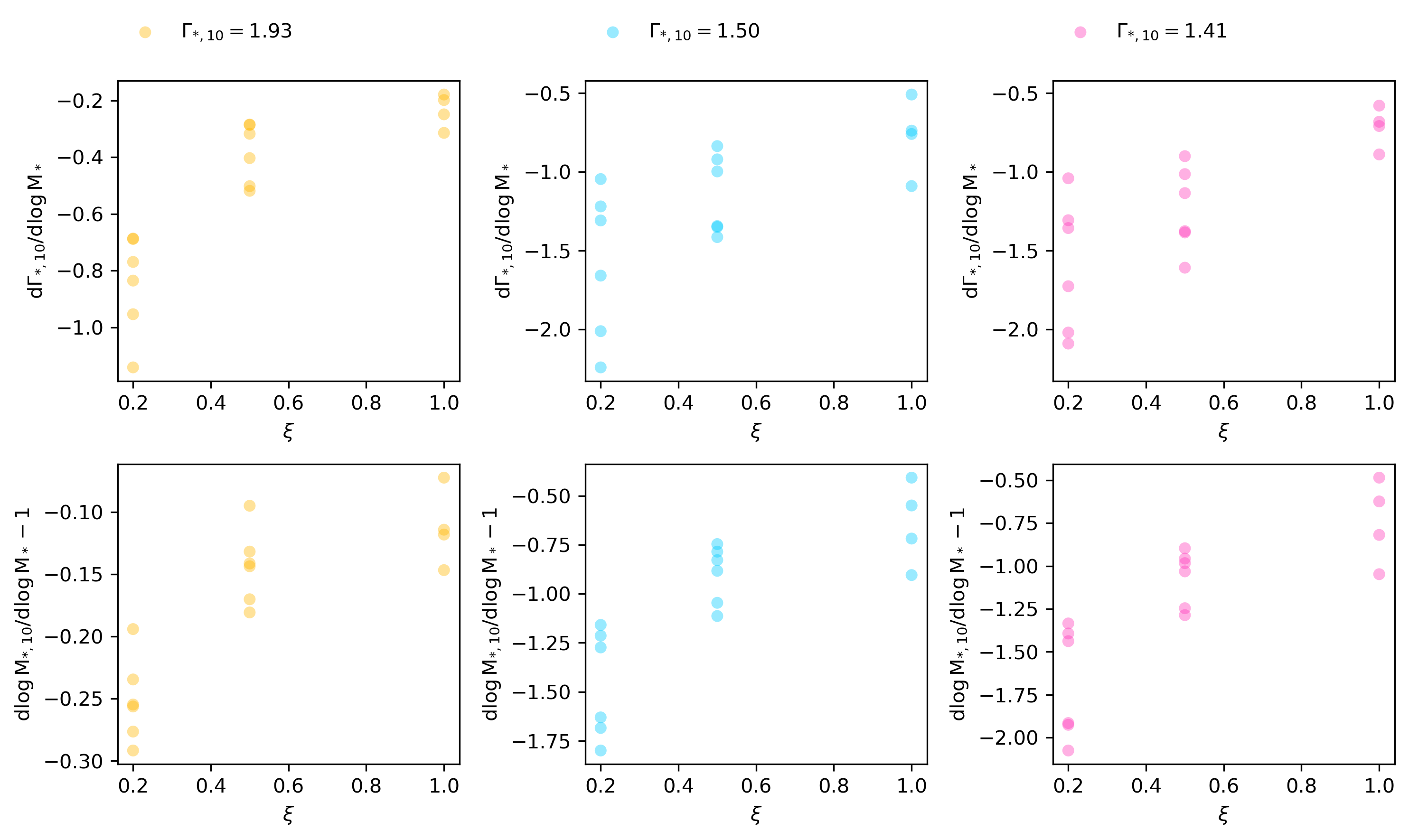}
    \caption{\label{fig:ddlogm}The first row shows the $\log M_*$ derivative of $\mathit{\Gamma_{*,10}}$ for each merger mass ratio $\xi$. Three different columns represent three different sets of mock galaxies with different $\mathit{\Gamma_{*,10}}$. The second row shows the $\log M_*$ derivative of $\log M_{*,10}$, subtracted by one to eliminate the effect of a homogeneous mass accretion. Each data point represents a particular run of simulations. For merger mass ratio $\xi=  1$, two sets of simulations with two runs in each set are shown. For the other two $\xi$, three sets of simulations are shown. Note in some cases two point overlap, hence the number of visible data points is smaller than the number of runs. } 
\end{figure*}
\par The next step to generate a mock observation is to give a physical unit to the simulation data. \cite{nipoti2009} used code units for length. Only after we specify a physical value for this length unit can we calculate the $\mathit{\Gamma_{*,10}}$ for our mock galaxies. We refer to such operation as rescaling the simulation data.  In order to understand what is the impact of mergers on the growth of $M_{*,10}$ and $\mathit{\Gamma_{*,10}}$, we needed a series of mock galaxies with different initial $M_{*,10}$ and $\mathit{\Gamma_{*,10}}$. Therefore, we needed to rescale the simulation data several times to obtain a series of mock galaxies. If we set the length unit to $1 ~\rm{kpc}$, the effective radius of the main galaxy would be $R_e = 1.22 ~\rm{kpc}$ while the $\mathit{\Gamma_{*,10}} \approx 1.93$. We then rescaled the simulation data, generating two additional sets of mock galaxies with effective radius $R_e = 8.5~\rm{kpc}$ and $R_e = 11.5~\rm{kpc}$ with $\mathit{\Gamma_{*,10}} \approx 1.50$ and $\mathit{\Gamma_{*,10}} \approx 1.44 $ respectively. The reason for choosing these specific values is that we want to let the $\mathit{\Gamma_{*,10}}$ value of the rescaled mock galaxies lay in the gray band in Fig.~\ref{fig:mg_relation} while having a moderate separation between each other, to make the following analyses more convenient. 
\par We then measured the evolution in $\log M_{*,10}$ and $\mathit{\Gamma_{*,10}}$ for the mock galaxies. Fig.~\ref{fig:ddlogm} shows the derivatives of $\log M_{*,10}$ and $\mathit{\Gamma_{*,10}}$ with respect to $\log M_*$ with three different sized ETGs due to mergers as a function of merger mass ratio $\xi$, while Fig.~\ref{fig:sim_result} shows $\Delta \mathit{\Gamma_{*,10}}^{\text{total}}$ versus $\Delta \log M_{*,10}^{\text{total}}$ for different merger mass ratios $\xi$. In these two figures, each data point correspond to one particular run of the simulations. The first row of Fig.~\ref{fig:ddlogm}, which shows the $\log M_*$ derivative of $\mathit{\Gamma_{*,10}}$ illustrates that mergers always decrease the $\mathit{\Gamma_{*,10}}$. The decrease can be explained by a scenario in which the stellar density profile is flattened by mergers, hence producing a shallower density slope. However, it is not straightforward to confirm this argument from the behavior of $M_{*,10}$, as the evolution in $M_{*,10}$ due to mergers is a bit complicated and can be decomposed in two parts. The first part is induced by the increase in total stellar mass, while the second part is due to the change in density profile. Nevertheless, the signal of the first part can be approximated by a homogeneous mass accretion, for which $\mathrm{d} \log M_{*,10} / \mathrm{d} \log M_* = 1$. Therefore, the net effect of a merger on the density profile can be captured by $\mathrm{d} \log M_{*,10} / \mathrm{d} \log M_* - 1 $, which is illustrated in the bottom row of Fig.~\ref{fig:ddlogm}. The negative value is evidence of a scenario in which the density profile is indeed flattened, as the mass of the inner region will outflow to the outer region during the flattening process, resulting in a decrease in $M_{*,10}$. A similar feature is also revealed in the work of~\cite{hilz_how_2013}, who performed N-body simulations of sequences of mergers, and showed that the slope of the stellar surface mass density profile becomes shallower after mergers (see their Figure 3). 
\par We can also conclude that for the same amount of mass that galaxies accrete via mergers, a smaller merger mass ratio $\xi$ results in a flatter stellar density profile. Both $\mathrm{d} \mathit{\Gamma_{*,10}} / \mathrm{d} \log M_*$ and $\mathrm{d} \log M_{*,10} / \mathrm{d} \log M_* - 1$ in Fig.~\ref{fig:ddlogm} show a larger absolute value for lower $\xi$. In addition, for the same merger mass ratio $\xi$, the change in density profile is more significant for galaxies that are larger in $R_e$ and smaller in $\mathit{\Gamma_{*,10}}$, as shown in Fig.~\ref{fig:sim_result}. We can observe that the absolute value of $\Delta \mathit{\Gamma_{*,10}}^\text{total}$ of $\mathit{\Gamma_{*,10}} \approx 1.93$ galaxies is always smaller than the other two sets of galaxies. In conclusion, the effect of mergers is to flatten the stellar density profile, while the flattening is stronger for lower $\xi$ mergers. 
\subsection{Mass growth}\label{sec:toy_model}
The evolution of the \gmr~depends on the mass growth of ETGs. With the help of the binary merger simulations, we related the fractional mass growth $f_M$ with $\zeta_\mu \text{ and } \zeta_\beta$ within redshift interval $0.17 \leq z \leq 0.37$ in the context of dry mergers.
\par As a first step, we considered a special condition in which all the mass growth is induced by mergers of a single mass ratio $\xi$, in particular $\xi = 0.2, 0.5, 1.0$ mergers. To achieve the goal of connecting $f_M$ to $\zeta_\mu$ and $\zeta_\beta$, we started from the fractional mass growth $f_M$, calculated the number of mergers $N_{merger}$ according to $f_M$, predicted the \gmr~after $N_{merger}$ merger events, and then calculated the resulting $\zeta_\mu$ and $\zeta_\beta$.
\par To calculate the number of merger events $N_{merger}$ in the first step, we simply divided $f_M$ by $\xi$: 
\begin{equation}
    N_{merger} \approx \frac{f_M}{\xi}.
\end{equation}
We note here that the number $N_{merger}$ is not a value that describes individual galaxies, but an average number of the entire galaxy population. Therefore $N_{merger}$ can in principle take any positive value, to be used to predict the average evolution of $M_{*,10}$ and $\mathit{\Gamma_{*,10}}$ for the galaxy population in the following step.
\par
\begin{figure*}
    \centering
    \includegraphics[width=\linewidth]{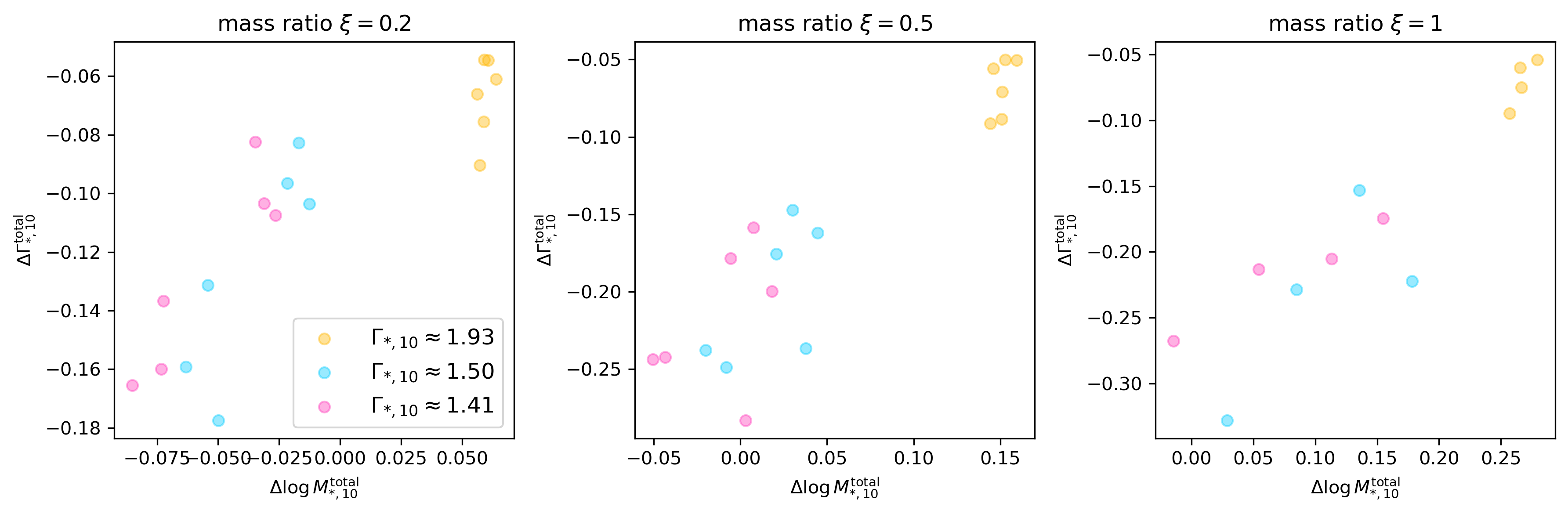}
    \caption{ Three panels show the change in $logM_{*,10}$ and $\mathit{\Gamma_{*,10}}$ induced by mergers. The three different colors represent different $\mathit{\Gamma_{*,10}}$ of the initial main galaxy. Each data point represents a particular run of simulations. For merger mass ratio $\xi=  1$, two sets of simulations with two runs in each set are shown. For the other two $\xi$, three sets of simulations are shown. Note in some cases two point overlap, hence the number of visible data points is smaller than the number of runs. }
    \label{fig:sim_result}
\end{figure*}

\begin{figure}
    \centering
    \includegraphics[width=\linewidth]{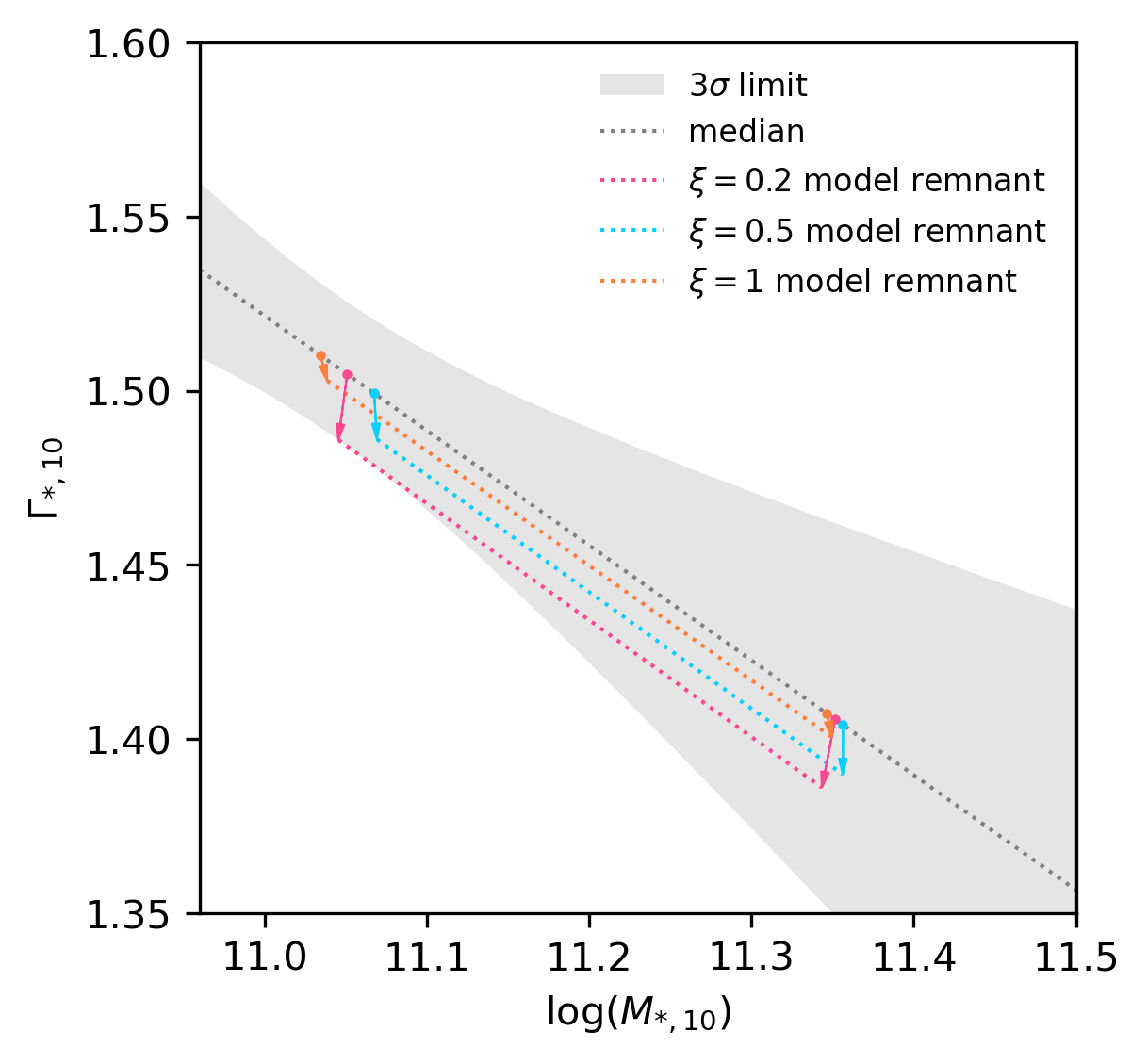}
    \caption{The evolution of the \gmr~due to different types of merger toy models. The gray band is the same as in Fig.~\ref{fig:mg_relation}, representing a upper limit of the evolution of the \gmr. We start from the median \gmr~of the gray band (shown in gray dashed line). We used the fitting formula of EMERGE \citep{2018moster} to estimate the number of mergers, thus calculating $\zeta_\mu$ and $\zeta_\beta$. The three sets of dots that lie on the starting \gmr~represent three different merger models, while the arrows show the change of these dots and the resulting \gmr.}
    \label{fig:toy}
\end{figure}

 The second step, predicting the \gmr~after $N_{merger}$ merger events, requires more effort. First we chose an initial \gmr~as the starting point. We set the starting \gmr~to be the median value of the gray band in Fig.~\ref{fig:mg_relation}. We located the initial state of mock galaxies with two rescaling settings, $\mathit{\Gamma_{*,10}} \approx 1.50$ and $\mathit{\Gamma_{*,10}} \approx 1.41$, on the starting \gmr. We show the initial state of mock galaxies in Fig.~\ref{fig:toy}. The three different colors mean three different $\xi$ mergers that the mock galaxies would have undergone. Due to different detailed settings in the simulation of those three $\xi$ mergers \citep[see table 1 of ][]{sonnenfeld2014}, the positions of mock galaxies in $\mathit{\Gamma_{*,10}}-M_{*,10}$ space have a slight offset with each other. Then we calculated how much do $M_{*,10}$ and $\mathit{\Gamma_{*,10}}$ of these two sets of rescaled galaxies evolve in the context of three different $\xi$ mergers, utilizing the simulation result and $N_{merger}$:
\begin{equation}
    \begin{aligned}
    \Delta M_{*,10} &= N_{merger} \times \Delta M_{*,10}^{\text{total}}\left(\xi \right) \\
    \Delta \mathit{\Gamma_{*,10}} &= N_{merger} \times \Delta \mathit{\Gamma_{*,10}}^{\text{total}}\left(\xi \right).
    \end{aligned}
\end{equation}
Here $\Delta M_{*,10}^{\text{total}}$ and $\Delta \mathit{\Gamma_{*,10}}^{\text{total}}$ are the median values for each group of simulation data with different merger mass ratios for an entire merger event, which are shown in Fig.~\ref{fig:sim_result}. Adding the growth to the initial state, we obtained the final state of our mock galaxies, then simply connecting the final state of two groups of mock galaxies by dotted lines gives the resulting \gmr. Fig.~\ref{fig:toy} is a illustration of how the \gmr~is evolved. In this figure, the value of $\Delta M_{*,10}$ and $\Delta \mathit{\Gamma_{*,10}}$ is calculated based on the fractional mass growth $f_M = 3.2\%$ from $z = 0.37$ to $z = 0.17$ that is obtained from \cite{2018moster}, which is further discussed in Sect.~\ref{sec:comparison}.
\par
\begin{figure}
    \centering
    \includegraphics[width=\linewidth]{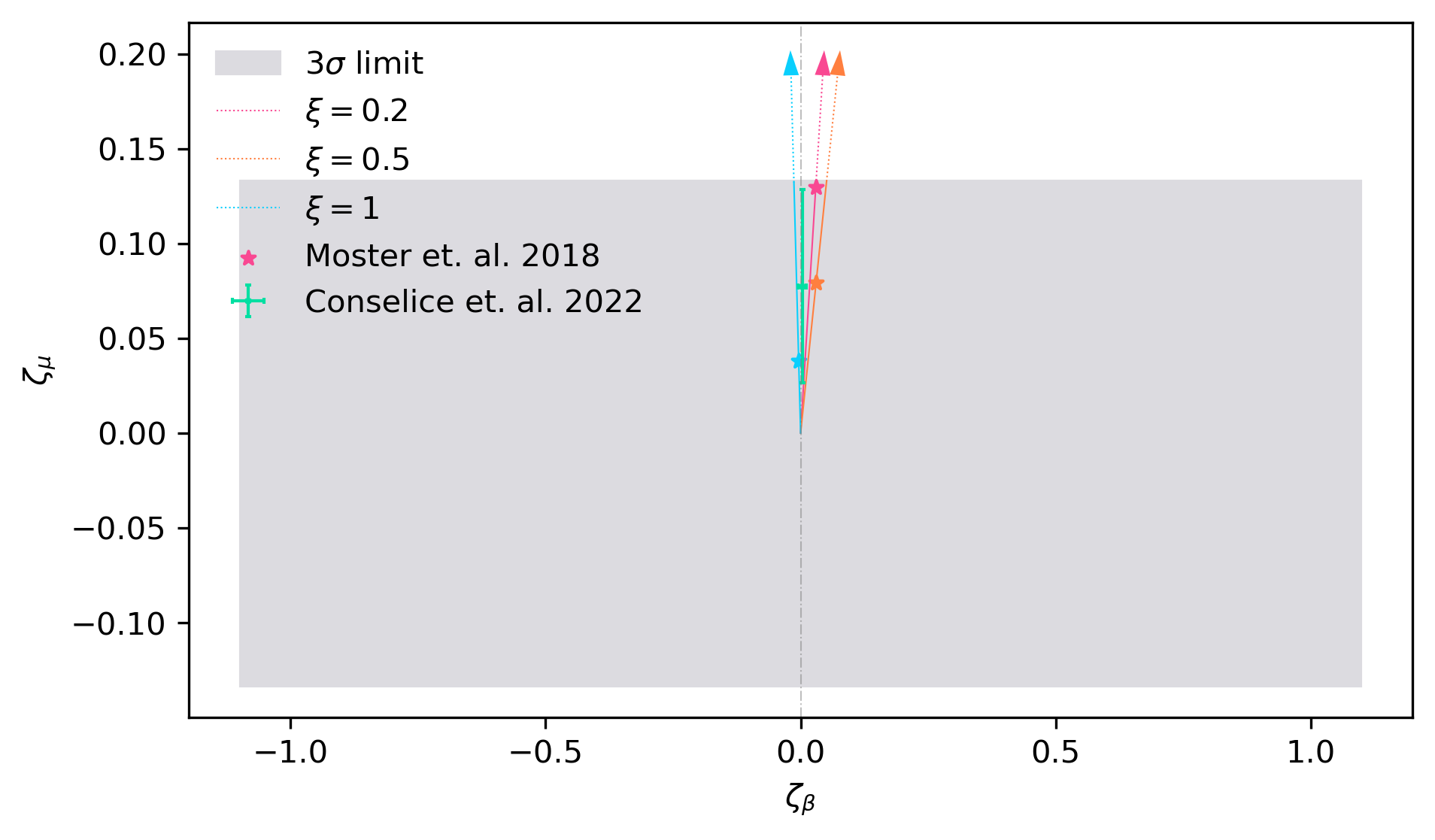}
    \caption{The upper limit of the evolution of the \gmr~ in $\zeta_\mu - \zeta_\beta$ space. The gray filled contour represents the estimate of the allowed evolution in the slope and normalization of the \gmr. The star symbol shows the prediction of $\xi = 0.2$ (in pink), $\xi = 0.5$ (in orange), and $\xi = 1$ (in blue) mergers with the merger number obtained from \cite{2018moster}. The lines with the arrow point in the direction of how $\zeta_\mu$ and $\zeta_\beta$ would evolve if the number of merger events increases. We also obtained the mass growth, in particular $f_M^{major}$ and $f_M^{minor}$, from \cite{conselice2022}. We sum up the evolution of the \gmr~due to major and minor mergers to get the resulting $\zeta_\mu$ and $\zeta_\beta$, and shows the result with the green error bar.  
    }
    \label{fig:conservative_contour}
\end{figure}

 The third step is to calculate the $\zeta_\mu$ and $\zeta_\beta$. We can obtain the change in normalization and slope by comparing the initial and resulting \gmr~given by the second step. By specifying a certain redshift interval, we can then calculate the redshift derivatives of the normalization and slope of the \gmr: $\zeta_\mu$ and $\zeta_\beta$.
For instance, applying this approach to the initial gray line with the three colored lines in Fig.~\ref{fig:toy}, we obtained the $\zeta_\mu$ and $\zeta_\beta$ for each $\xi$ merger given the $f_M$ in redshift range $0.17 \leq z \leq 0.37$ from \cite{2018moster}. The results are shown in Fig.~\ref{fig:conservative_contour} with star symbols as an illustration. 
\par In order to achieve the goal of finding a relation between $f_M$ and $\zeta_\mu, \zeta_\beta$ in the redshift interval $0.17 \leq z \leq 0.37$, we repeated the above three steps several times, with each iteration using a larger fractional mass growth $f_M$. The three colored lines in Fig.~\ref{fig:conservative_contour} show how $\zeta_\mu$ and $\zeta_\beta$ evolve with $f_M$, with the arrow pointing to the direction in which $f_M$ increases. For different merger mass ratio, the direction of the arrow is different. All mergers decrease the normalization of the \gmr, while the slope becomes steeper for $\xi = 0.2$ and $\xi = 0.5$ mergers, but becomes shallower for $\xi = 1.0$ mergers. 
\par Now we have successfully connected $\zeta_\mu$ and $\zeta_\beta$ to $f_M$ in $0.17 \leq z \leq 0.37$ for the condition that only mergers with given merger mass ratio $\xi$ contribute to the mass growth of ETGs. In reality, however, the mass growth should be induced by a multitude of mergers involving different mass ratio mergers. The total mass growth $f_M^{tot}$ is the sum of the mass growth $f_M(\xi)$ induced by different $\xi$ mergers 
\begin{equation}\label{eq:sumfm}
    f_M^{tot} = \int f_M(\xi) d\xi.
\end{equation} 
In this case, the first and second steps described above need some modification. We need to calculate $N_{merger}(\xi)$ and thus obtain $\Delta M_{*,10}(\xi)$ and $\Delta \mathit{\Gamma_{*,10}}(\xi)$ for each $\xi$, and sum up $\Delta M_{*,10}(\xi)$ and $\Delta \mathit{\Gamma_{*,10}}(\xi)$ to get the total $\Delta M_{*,10}$ and $\Delta \mathit{\Gamma_{*,10}}$. Hence, an accurate $f_M(\xi)$ is required to find the accurate relation between $f_M^{tot}$ and $\zeta_\mu, \zeta_\beta$. The different directions of the three arrows in Fig.~\ref{fig:conservative_contour} suggests that if we can measure $\zeta_\mu$ and $\zeta_\beta$ precisely, we might be able to infer the fractional mass growth $f_M(\xi)$ for different $\xi$ mergers. However, our observation is currently limited by systematics, thus not sufficiently precise to infer $f_M(\xi)$.
\par Nevertheless, our initial goal, which is to explore to what extent the evolution of the \gmr~in redshift range $0.17 \leq z \leq 0.37$ can be explained by dry mergers, can still be achieved. We can exploit the fact that smaller mass ratio $\xi$ mergers can result in a more significant evolution in the \gmr, which is reflected by the three star symbols in Fig.~\ref{fig:conservative_contour}. The three symbols have the same fractional mass growth, but the $\zeta_\mu$ and $\zeta_\beta$ of the $\xi = 0.2$ merger is the largest. In other words, mergers with larger $\xi$ must induce more mass growth of ETGs to achieve the same degree of evolution in the \gmr~as mergers with smaller $\xi$ values. Therefore, we can use the largest $\xi$, which is $\xi = 1$, to place an upper limit on the fractional mass growth $f_M$. Using the toy model for a single mass ratio $\xi = 1$ mergers, the maximum fractional mass growth $f_M$ in the redshift range $0.17 < z < 0.37$ turns out to be $11.2\%$. This is the $f_M$ boundary where the dry merger scenario works. If the fractional mass growth $f_M$ of ETGs in redshift range $0.17 \leq z \leq 0.37$ is smaller than $11.2\%$, or $\Delta \log M_* / \Delta t \leq 0.022 ~\rm{dex/Gyr}$, then the dry merger scenario is consistent with our observational bounds on the evolution of the \gmr. 


\section{Discussion}\label{sec:4}
\subsection{Comparison with the literature}\label{sec:comparison}
In our observations, we cannot find evidence for an evolution of the \gmr. 
This is qualitatively consistent with \cite{Huang2018}, whose Figure 6 shows that the inner mass profile of massive galaxies at redshift range $0.25 \leq z \leq 0.50$ are similar to those of $z \sim 0$ massive galaxies. 
\par We made a comparison between our upper limit on the fractional mass growth with works from the literature. We first compared $f_M$ with an empirical model, EMERGE \citep{2018moster}. EMERGE is a semi-analytic model that is designed to populate mock galaxies inside simulated dark matter halos, thus obtaining predictions about the galaxy formation and evolution from the mock galaxies. For the EMERGE model, we obtained the fractional mass growth $f_M$ in the redshift range $0.17 \leq z \leq 0.37$ from the fraction of stellar mass accreted to the main galaxy as a function of redshift, $f_{acc}$, with respect to $z=0$. This is given by Eq.24 in \cite{2018moster}:
\begin{equation}
    f_{acc}(z) = 0.55\exp \left[-1.09(z+1)\right].
\end{equation}
Hence $f_M$ can be calculated by substituting the lower and upper redshift of our sample 
\begin{equation}\label{eq:fm}
    f_M = \frac{1-f_{acc}(0)+f_{acc}(0.17)}{1-f_{acc}(0)+f_{acc}(0.37)} - 1 = 3.2\%.
\end{equation}
This value is smaller than the maximum allowed fractional mass growth $f_M = 11.2\%$ that we obtained from the toy model. We further checked the induced $\zeta_\mu$ and $\zeta_\beta$, using the procedure introduced in Sect.~\ref{sec:toy_model} for three different $\xi$ mergers. The three star symbols in Fig.~\ref{fig:conservative_contour}, representing the result position in $\zeta_\mu - \zeta_\beta$ space, all lies inside the gray contour. Therefore, we consider the EMERGE result as consistent with our observation. 
\par Next, we compared our results with the observation of \cite{conselice2022}. \cite{conselice2022} defined major mergers as mergers with mass ratio $\xi \geq 0.25$, and minor mergers as mergers with mass ratio $0.1 \leq \xi \leq 0.25$. They selected galaxies from the REFINE survey in the redshift range $0 < z < 3$ and separated them into different redshift bins, then measured the number of merger events in each redshift bin and inferred the amount of accreted mass as a function of redshift. For comparison, we picked the redshift bin $0.2 < z < 0.5$, as it is the closest to our redshift range $0.17 \leq z \leq 0.37$. According to \cite{conselice2022}, the stellar mass of the main galaxies, which will merge with their nearby lower mass companions later, is $\log M_* = 11.3 \pm 0.1 $ in this redshift bin, and the mass accreted via major and minor mergers is $\log M_{*,acc}^{major} = 9.9 \pm 0.2$ and $\log M_{*,acc}^{minor} = 9.2 \pm 0.3$ respectively \citep[see table 2 in ][]{conselice2022}. Assuming that the accretion rate is constant during $0.2 < z < 0.5$, we calculated the total accreted mass $M_{*,acc} = M_{*,acc}^{major} + M_{*,acc}^{minor}$ in our focused redshift range $0.17 \leq z \leq 0.37$, and then calculated the fractional mass growth $f_M = M_{*,acc} / M_*  = 3.5 ^{+3.8}_{-1.8} \%$. Further, we used the $\xi = 0.2$ merger scenario to evolve the \gmr~according to $M_{*,acc}^{minor}$, and the $\xi = 1$ merger scenario to evolve the \gmr~according to $M_{*,acc}^{major}$. The result is shown in Fig.~\ref{fig:conservative_contour} with the green point, with the error bar representing the uncertainty. The green point lies inside the gray contour, indicating that the result of \cite{conselice2022} is also consistent with our observations.
\par Although the fractional mass growth given by EMERGE and \cite{conselice2022} observations is similar, the induced $\zeta_\mu$ and $\zeta_\beta$ are different, as shown in Fig.~\ref{fig:conservative_contour}. We believe that we still have the potential to distinguish different merger scenarios, as long as we have sufficiently precise measurement. To precisely measure $\zeta_\mu$ and $\zeta_\beta$, we need to overcome the limitation of systematics. 
The pink star symbol in Fig.~\ref{fig:conservative_contour} shows the induced $\zeta_\mu$ and $\zeta_\beta$ of the $f_M$ from EMERGE model. Although this symbol correspond to the most extreme case in which all mergers have the smallest mass ratio $\xi = 0.2$, it still lies inside the gray contour, which indicates that the systematic inside the data might be larger in comparison with the intrinsic evolution of the \gmr. We hence discussed the possible source of systematics in the following Sect.~\ref{sec:systematic}. 
\par Nevertheless, we believe that if we apply our method to a larger redshift baseline, we can potentially detect the real signal of evolution in the \gmr, thus achieve a better understanding of the late growth of ETGs. To make a rough estimate, we utilized the $\zeta_\mu$ and $\zeta_\beta$ derived from the result of \cite{conselice2022}. We find that if we can explore to a higher redshift of $z = 0.54$, i.e. a time interval $\Delta t = 3.2~\mathrm{Gyr}$, the \gmr~will evolve outside the gray contour. Hence, we believe that if we can focus on a redshift range with time interval longer than $\Delta t \geq 3.2~\mathrm{Gyr}$, the signal of the evolution of the \gmr~will not be buried under the systematics.
\subsection{Possible source of systematics}\label{sec:systematic}
In our observations, we chose to be conservative and hence found the gray band that serves as the upper limit of the evolution of the \gmr. The \gmr~could have evolved its slope and normalization in the redshift range $0.17 \leq z \leq 0.37$, but should be constrained inside the gray band. We suspect that quantities that were derived from photometry could be 
affected by systematics, but we lack robust ways to quantitatively estimate and eliminate them.

One possible source of bias is the stellar population synthesis model, on which the measurement of the stellar mass-to-light ratio of our galaxies are based.
Stellar population synthesis fits to broadband imaging data are prone to systematic errors \citep[see e.g.][]{Conroy2013}, and these errors can be redshift-dependent.
For instance, \citet{Paulino2022} found that the stellar masses of $z=0.5$ galaxies are underestimated by $0.1-0.2$~dex, with respect to $z=0$ counterparts.
Nevertheless, we suspect that such biases are highly dependent on the details of the stellar population modeling method and data used, therefore we refrain from applying a correction to our results based on their analysis.

\begin{figure}
    \centering
    \includegraphics[width=0.95\linewidth]{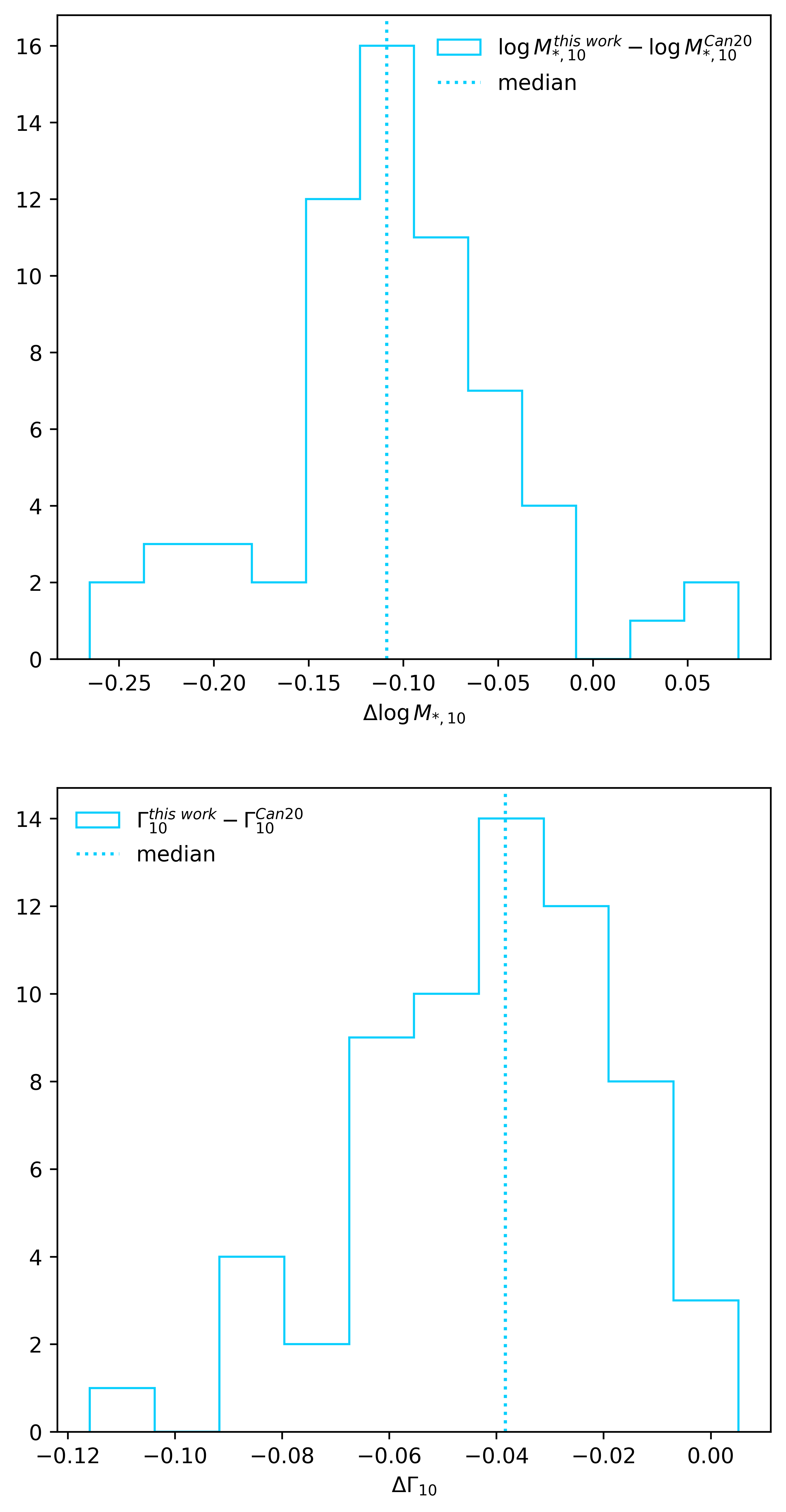}
    \caption{The difference between the derived $M_{*,10}$ (upper panel) and $\Gamma_{*,10}$ (bottom panel) using different S\'{e}rsic parameters ($R_e$ and $n$). The superscript `this work' means the $\reff$ and $n$ are from this work, while the superscript `Can20' means the $\reff$ and $n$ are from \cite{carlo2020}. The dashed line represents the median value of the difference.}
    \label{fig:systematic}
\end{figure}

Another issue that might induce a systematic is the presence of color gradients. Galaxies are typically bluer in the outer regions \citep[e.g.][]{Suess2019a, Suess2019b, Suess2020}, thus they look more extended when observed at shorter wavelengths. We compared our observations with those of \cite{carlo2020}, computing $M_{*,10}$ and $\mathit{\Gamma_{*,10}}$ of 63 overlapping galaxies from their S\'{e}rsic model fits. We found some discrepancy between the two datasets, as shown in Fig.~\ref{fig:systematic}. Both $\log M_{*,10}$ and $\mathit{\Gamma_{*,10}}$ are systematically smaller in our sample than in \cite{carlo2020}, with a median discrepancy of $0.1$ dex and $0.04$ respectively. The depths of both surveys are sufficient to capture the full extent of galaxy light within a radius of $10~\rm{kpc}$. For our highest-redshift subsample ($0.39 < z < 0.40$), the typical surface brightness at $10~\rm{kpc}$ is approximately $24~\mathrm{mag/arcsec^2}$. This is well above the detection threshold of the shallower KiDS survey, which is typically $26~\mathrm{mag/arcsec^2}$ measured from galaxies shown in Fig.~\ref{fig:surface_brightness}. Therefore, the main difference between these two measurements is that they use observed data within different ranges of wavelength. \cite{carlo2020} use $g, r, i, z, ~\text{and}~ y$-band HSC SSP data, where the $i$-band data has the best S/N ratio. Hence we believe the $i$-band takes the major role in determining the stellar mass as well as the surface brightness profile. Meanwhile, in our measurement we use the data in a fixed wavelength $3000 - 10000 $ \AA~to measure the stellar mass and only use the data in $r$-band, which has a shorter wavelength than the $i$-band, to measure the S\'{e}rsic parameters. The discrepancy in $\mathit{\Gamma_{*,10}}$ measurement can be qualitatively explained by the existence of the color gradient, as the smaller value of $\mathit{\Gamma_{*,10}}$ in our sample is consistent with the trend that galaxies look more extended in bluer filter band. The choice that we have made to ignore the color gradient in our sample therefore introduce a potential systematic bias on $M_{*,10}$ and $\mathit{\Gamma_{*,10}}$ measurement. 

This bias could be alleviated by obtaining spatially-resolved stellar population synthesis models, though at the expense of a significant increase in the complexity of our analysis.
We defer such an attempt to future work.
\par Finally, another important systematic effect is the progenitor bias \citep{van_dokkum_1996,saglia2010, carolloNEWLYQUENCHEDGALAXIES2013,fagioliMinorMergersProgenitor2016}. The population of quiescent galaxies will be continuously augmented by the addition of recently quenched ex-star-forming galaxies, thus also leaving imprints on the average \gmr. 
One solution to remove this bias could be tracing the evolution track for both star-forming galaxies and quiescent galaxies. Progenitor bias does not necessarily impact our measurements, but it does complicate their interpretation. In order to take progenitor bias into account, we would need to analyze star-forming galaxies in addition to the ETGs. However, the transition from star-forming to quiescent galaxy changes the morphology of galaxies in a non-trivial way, so it is not clear how to map objects from one population to the other. Moreover, comparing stellar mass measurements of galaxy populations with very different star formation history can be subject to biases. 
Another possible solution is to focus on ETGs only, and trace the descendants of higher redshift ETGs with some proxies, for instance the $D_n4000$ index \citep{zahid2019,hamadouche2022,Damjanov2023}. In principle, we can use stellar population synthesis models to predict the evolution of the $D_n4000$ index of quiescent galaxies, and apply a redshift-dependent $D_n4000$ cut to follow galaxies along their evolutionary track. However, further investigation is required to validate the robustness of such approach to choices of the underlying stellar population synthesis model. 
We thus leave this analysis to future work.

\section{Conclusion}\label{sec:5}
In our work, we provided a new perspective to investigate the evolution of quiescent galaxies. Instead of trying to obtain information on stellar mass and effective radius from the entire light of one galaxy, we switched our attention to the inner region. In particular, we focused on a fixed physical aperture $10kpc$ and measured the stellar mass and stellar mass-weighted projected surface brightness slope $\mathit{\Gamma_{*,10}}$ enclosed in this aperture, namely the $10~\rm{kpc}$ collar. 
\par First, combining KiDS imaging with GAMA spectroscopy, we measured the \gmr~in four different redshift bins with a Bayesian hierarchical model, while correcting for the Eddington bias. Due to potential uncertainties and systematic effects, current observations are not sufficient to detect an evolution in the \gmr. We put an upper limit on the parameters $\zeta_\mu$ and $\zeta_\beta$, which are the redshift derivative of the normalization and the slope of the \gmr. The values of the upper limits are $|\zeta_\mu| \leq 0.13$ and $\left|\zeta_\beta\right| \leq 1.10$.
\par For the interpretation of these observations we focused on dissipationless mergers, which is a well-agreed mechanism that could drive the growth of galaxies. In order to understand how mergers evolve quiescent galaxies in $\mathit{\Gamma_{*,10}} - M_{*,10}$ space, we utilized a collection of dissipationless binary merger simulations. We analyzed the difference between pre-merger mock galaxies and post-merger galaxies and drew the following conclusions: 
\begin{enumerate}[(i.)]
    \item Mergers flatten the density profile, thus result in a decrease in $\mathit{\Gamma_{*,10}}$. This is reflected by the fact that all post-merger mock galaxies have a smaller $\Gamma_{*,10}$ compared with their pre-merger counterparts. The larger decrease in $\mathit{\Gamma_{*,10}}$ for lower $\xi$ mergers indicates that the flattening is stronger for smaller mass ratio mergers.
    \item In most cases, mergers induce an increase in $M_{*,10}$. However, $M_{*,10}$ is observed to decrease when galaxies with low $\Gamma_{*,10}$ undergo a low-$\xi$ merger. This indicates that the effect of minor mergers is to redistribute stellar mass towards the outer regions. 
\end{enumerate}
\par After obtaining the knowledge on how mergers evolve galaxies in the $M_{*,10} - \mathit{\Gamma_{*,10}}$ space, we established a toy model to explore to what extent can a dry merger scenario explain the possible evolution of the \gmr. We found that the effect of mergers on the \gmr~is different depending on the merger mass ratio. The normalization of the \gmr~always decreases after mergers, while the slope of the \gmr~becomes either shallower or steeper, depending on the merger mass ratio. We then used the upper limit on $\zeta_\mu$ and $\zeta_\beta$ to find the maximum fractional mass growth $f_M = 11.2\%$ of ETGs during $0.17 \leq z \leq 0.37$, below which the dry merger scenario is consistent with our observations. We compared our maximum $f_M$ with the empirical model EMERGE \cite{2018moster} and the observations of \cite{conselice2022}, and found that our work is consistent with both of them.
\par Besides, we discussed the possible source of biases or systematics in our observations. We found that the measurement of both $M_{*,10}$ and $\mathit{\Gamma_{*,10}}$ seems to have a dependence on the wavelength used in analysis, which might be evidence that the color gradient of galaxies should not be ignored. Another potential source of bias could be the contamination from newly quenched star-forming galaxies. In existence of these systematics, we may need a larger redshift baseline, particularly $\Delta z \geq 0.45$, to observe a signal of the evolution of the \gmr. More efforts are needed to either extend the redshift baseline or eliminate the  possible systematics, thus help us gain a better understanding of the post-quenching evolution of ETGs.   
\par In conclusion, we provided a new method to investigate the evolution of quiescent galaxies and established a toy model based on N-body simulations to predict the impact of different mergers on the \gmr. When applied to data from the GAMA survey, our method is limited by systematics in the stellar mass estimate. Hopefully, the incoming surveys ,e.g. Euclid \citep{Euclid}, CSST \citep{csst} could widen the survey coverage while deepening the survey's limiting flux. We believe we would be able to overcome the limitation and give a more precise conclusion on the evolution of quiescent galaxies in near future. 
\begin{acknowledgements}
This work was supported by the National Key R\&D Program of C'hina (No. 2023YFA1607800, 2023YFA1607802), and made use of the Gravity Supercomputer at the Department of Astronomy, Shanghai Jiao Tong University.
\end{acknowledgements}

\bibliographystyle{aa}
\bibliography{references}

\begin{thebibliography}{73}
\expandafter\ifx\csname natexlab\endcsname\relax\def\natexlab#1{#1}\fi

\bibitem[{{Auger} {et~al.}(2010){Auger}, {Treu}, {Bolton}, {Gavazzi}, {Koopmans}, {Marshall}, {Moustakas}, \& {Burles}}]{auger2010}
{Auger}, M.~W., {Treu}, T., {Bolton}, A.~S., {et~al.} 2010, \apj, 724, 511

\bibitem[{{Baldry} {et~al.}(2010){Baldry}, {Robotham}, {Hill}, {Driver}, {Liske}, {Norberg}, {Bamford}, {Hopkins}, {Loveday}, {Peacock}, {Cameron}, {Croom}, {Cross}, {Doyle}, {Dye}, {Frenk}, {Jones}, {van Kampen}, {Kelvin}, {Nichol}, {Parkinson}, {Popescu}, {Prescott}, {Sharp}, {Sutherland}, {Thomas}, \& {Tuffs}}]{GAMA1}
{Baldry}, I.~K., {Robotham}, A.~S.~G., {Hill}, D.~T., {et~al.} 2010, \mnras, 404, 86

\bibitem[{{Balogh} {et~al.}(1999){Balogh}, {Morris}, {Yee}, {Carlberg}, \& {Ellingson}}]{Balogh99}
{Balogh}, M.~L., {Morris}, S.~L., {Yee}, H.~K.~C., {Carlberg}, R.~G., \& {Ellingson}, E. 1999, \apj, 527, 54

\bibitem[{{Belli} {et~al.}(2015){Belli}, {Newman}, \& {Ellis}}]{Belli_2015}
{Belli}, S., {Newman}, A.~B., \& {Ellis}, R.~S. 2015, \apj, 799, 206

\bibitem[{Bellstedt {et~al.}(2020)Bellstedt, Driver, Robotham, Davies, Bogue, Cook, Hashemizadeh, Koushan, Taylor, Thorne, Turner, \& Wright}]{bellstedt_galaxy_2020}
Bellstedt, S., Driver, S.~P., Robotham, A. S.~G., {et~al.} 2020, \mnras, 496, 3235

\bibitem[{Bruzual \& Charlot(2003)}]{bruzual_2003}
Bruzual, G. \& Charlot, S. 2003, \mnras, 344, 1000

\bibitem[{{Bundy} {et~al.}(2017){Bundy}, {Leauthaud}, {Saito}, {Maraston}, {Wake}, \& {Thomas}}]{Bundy2017}
{Bundy}, K., {Leauthaud}, A., {Saito}, S., {et~al.} 2017, \apj, 851, 34

\bibitem[{{Calzetti} {et~al.}(2000){Calzetti}, {Armus}, {Bohlin}, {Kinney}, {Koornneef}, \& {Storchi-Bergmann}}]{calzetti2000}
{Calzetti}, D., {Armus}, L., {Bohlin}, R.~C., {et~al.} 2000, \apj, 533, 682

\bibitem[{{Cannarozzo} {et~al.}(2020){Cannarozzo}, {Sonnenfeld}, \& {Nipoti}}]{carlo2020}
{Cannarozzo}, C., {Sonnenfeld}, A., \& {Nipoti}, C. 2020, \mnras, 498, 1101

\bibitem[{Carollo {et~al.}(2013)Carollo, Bschorr, Renzini, Lilly, Capak, Cibinel, Ilbert, Onodera, Scoville, Cameron, Mobasher, Sanders, \& Taniguchi}]{carolloNEWLYQUENCHEDGALAXIES2013}
Carollo, C.~M., Bschorr, T.~J., Renzini, A., {et~al.} 2013, \apj, 773, 112

\bibitem[{{Chabrier}(2003)}]{chabrier2003}
{Chabrier}, G. 2003, \pasp, 115, 763

\bibitem[{{Conroy}(2013)}]{Conroy2013}
{Conroy}, C. 2013, \araa, 51, 393

\bibitem[{{Conselice} {et~al.}(2022){Conselice}, {Mundy}, {Ferreira}, \& {Duncan}}]{conselice2022}
{Conselice}, C.~J., {Mundy}, C.~J., {Ferreira}, L., \& {Duncan}, K. 2022, \apj, 940, 168

\bibitem[{Daddi {et~al.}(2005)Daddi, Renzini, Pirzkal, Cimatti, Malhotra, Stiavelli, Xu, Pasquali, Rhoads, Brusa, {di Serego Alighieri}, Ferguson, Koekemoer, Moustakas, Panagia, \& Windhorst}]{daddiPassivelyEvolvingEarlyType2005}
Daddi, E., Renzini, A., Pirzkal, N., {et~al.} 2005, \apj, 626, 680

\bibitem[{{Damjanov} {et~al.}(2023){Damjanov}, {Sohn}, {Geller}, {Utsumi}, \& {Dell'Antonio}}]{Damjanov2023}
{Damjanov}, I., {Sohn}, J., {Geller}, M.~J., {Utsumi}, Y., \& {Dell'Antonio}, I. 2023, \apj, 943, 149

\bibitem[{Damjanov {et~al.}(2019)Damjanov, Zahid, Geller, Utsumi, Sohn, \& Souchereau}]{damjanov2019}
Damjanov, I., Zahid, H.~J., Geller, M.~J., {et~al.} 2019, \apj, 872, 91

\bibitem[{{de Jong} {et~al.}(2013){de Jong}, {Kuijken}, {Applegate}, {Begeman}, {Belikov}, {Blake}, {Bout}, {Boxhoorn}, {Buddelmeijer}, {Buddendiek}, {Cacciato}, {Capaccioli}, {Choi}, {Cordes}, {Covone}, {Dall'Ora}, {Edge}, {Erben}, {Franse}, {Getman}, {Grado}, {Harnois-Deraps}, {Helmich}, {Herbonnet}, {Heymans}, {Hildebrandt}, {Hoekstra}, {Huang}, {Irisarri}, {Joachimi}, {K{\"o}hlinger}, {Kitching}, {La Barbera}, {Lacerda}, {McFarland}, {Miller}, {Nakajima}, {Napolitano}, {Paolillo}, {Peacock}, {Pila-Diez}, {Puddu}, {Radovich}, {Rifatto}, {Schneider}, {Schrabback}, {Sifon}, {Sikkema}, {Simon}, {Sutherland}, {Tudorica}, {Valentijn}, {van der Burg}, {van Uitert}, {van Waerbeke}, {Velander}, {Verdoes Kleijn}, {Viola}, \& {Vriend}}]{deJong2013}
{de Jong}, J.~T.~A., {Kuijken}, K., {Applegate}, D., {et~al.} 2013, The Messenger, 154, 44

\bibitem[{{Dehnen}(1993)}]{Dehnen93}
{Dehnen}, W. 1993, \mnras, 265, 250

\bibitem[{Dekel \& Burkert(2014)}]{dekel_wet_2014}
Dekel, A. \& Burkert, A. 2014, \mnras, 438, 1870

\bibitem[{{D'Eugenio} {et~al.}(2023){D'Eugenio}, {van der Wel}, {Piotrowska}, {Bezanson}, {Taylor}, {van de Sande}, {Baker}, {Bell}, {Bellstedt}, {Bland-Hawthorn}, {Bluck}, {Brough}, {Bryant}, {Colless}, {Cortese}, {Croom}, {Derkenne}, {van Dokkum}, {Fisher}, {Foster}, {Gallazzi}, {de Graaff}, {Groves}, {van Houdt}, {del P. Lagos}, {Looser}, {Maiolino}, {Maseda}, {Mendel}, {Nersesian}, {Pacifici}, {Poci}, {Remus}, {Sweet}, {Thater}, {Tran}, {{\"U}bler}, {Valenzuela}, {Wisnioski}, \& {Zibetti}}]{deugenio2023}
{D'Eugenio}, F., {van der Wel}, A., {Piotrowska}, J.~M., {et~al.} 2023, \mnras, 525, 2789

\bibitem[{Dokkum \& Franx(1996)}]{van_dokkum_1996}
Dokkum, P. G.~v. \& Franx, M. 1996, \mnras, 281, 985

\bibitem[{{Driver} {et~al.}(2022){Driver}, {Bellstedt}, {Robotham}, {Baldry}, {Davies}, {Liske}, {Obreschkow}, {Taylor}, {Wright}, {Alpaslan}, {Bamford}, {Bauer}, {Bland-Hawthorn}, {Bilicki}, {Bravo}, {Brough}, {Casura}, {Cluver}, {Colless}, {Conselice}, {Croom}, {de Jong}, {D'Eugenio}, {De Propris}, {Dogruel}, {Drinkwater}, {Dvornik}, {Farrow}, {Frenk}, {Giblin}, {Graham}, {Grootes}, {Gunawardhana}, {Hashemizadeh}, {H{\"a}u{\ss}ler}, {Heymans}, {Hildebrandt}, {Holwerda}, {Hopkins}, {Jarrett}, {Heath Jones}, {Kelvin}, {Koushan}, {Kuijken}, {Lara-L{\'o}pez}, {Lange}, {L{\'o}pez-S{\'a}nchez}, {Loveday}, {Mahajan}, {Meyer}, {Moffett}, {Napolitano}, {Norberg}, {Owers}, {Radovich}, {Raouf}, {Peacock}, {Phillipps}, {Pimbblet}, {Popescu}, {Said}, {Sansom}, {Seibert}, {Sutherland}, {Thorne}, {Tuffs}, {Turner}, {van der Wel}, {van Kampen}, \& {Wilkins}}]{GAMAmain}
{Driver}, S.~P., {Bellstedt}, S., {Robotham}, A. S.~G., {et~al.} 2022, \mnras, 513, 439

\bibitem[{{D'Souza} {et~al.}(2014){D'Souza}, {Kauffman}, {Wang}, \& {Vegetti}}]{dsouza2014}
{D'Souza}, R., {Kauffman}, G., {Wang}, J., \& {Vegetti}, S. 2014, \mnras, 443, 1433

\bibitem[{{Eisenstein} {et~al.}(2023){Eisenstein}, {Willott}, {Alberts}, {Arribas}, {Bonaventura}, {Bunker}, {Cameron}, {Carniani}, {Charlot}, {Curtis-Lake}, {D'Eugenio}, {Endsley}, {Ferruit}, {Giardino}, {Hainline}, {Hausen}, {Jakobsen}, {Johnson}, {Maiolino}, {Rieke}, {Rieke}, {Rix}, {Robertson}, {Stark}, {Tacchella}, {Williams}, {Willmer}, {Baker}, {Baum}, {Bhatawdekar}, {Boyett}, {Chen}, {Chevallard}, {Circosta}, {Curti}, {Danhaive}, {DeCoursey}, {de Graaff}, {Dressler}, {Egami}, {Helton}, {Hviding}, {Ji}, {Jones}, {Kumari}, {L{\"u}tzgendorf}, {Laseter}, {Looser}, {Lyu}, {Maseda}, {Nelson}, {Parlanti}, {Perna}, {Pusk{\'a}s}, {Rawle}, {Rodr{\'\i}guez Del Pino}, {Sandles}, {Saxena}, {Scholtz}, {Sharpe}, {Shivaei}, {Silcock}, {Simmonds}, {Skarbinski}, {Smit}, {Stone}, {Suess}, {Sun}, {Tang}, {Topping}, {{\"U}bler}, {Villanueva}, {Wallace}, {Whitler}, {Witstok}, \& {Woodrum}}]{Eisenstein_JADES}
{Eisenstein}, D.~J., {Willott}, C., {Alberts}, S., {et~al.} 2023, arXiv e-prints, arXiv:2306.02465

\bibitem[{{Euclid Collaboration} {et~al.}(2022){Euclid Collaboration}, {Scaramella}, {Amiaux}, {Mellier}, {Burigana}, {Carvalho}, {Cuillandre}, {Da Silva}, {Derosa}, {Dinis}, {Maiorano}, {Maris}, {Tereno}, {Laureijs}, {Boenke}, {Buenadicha}, {Dupac}, {Gaspar Venancio}, {G{\'o}mez-{\'A}lvarez}, {Hoar}, {Lorenzo Alvarez}, {Racca}, {Saavedra-Criado}, {Schwartz}, {Vavrek}, {Schirmer}, {Aussel}, {Azzollini}, {Cardone}, {Cropper}, {Ealet}, {Garilli}, {Gillard}, {Granett}, {Guzzo}, {Hoekstra}, {Jahnke}, {Kitching}, {Maciaszek}, {Meneghetti}, {Miller}, {Nakajima}, {Niemi}, {Pasian}, {Percival}, {Pottinger}, {Sauvage}, {Scodeggio}, {Wachter}, {Zacchei}, {Aghanim}, {Amara}, {Auphan}, {Auricchio}, {Awan}, {Balestra}, {Bender}, {Bodendorf}, {Bonino}, {Branchini}, {Brau-Nogue}, {Brescia}, {Candini}, {Capobianco}, {Carbone}, {Carlberg}, {Carretero}, {Casas}, {Castander}, {Castellano}, {Cavuoti}, {Cimatti}, {Cledassou}, {Congedo}, {Conselice}, {Conversi}, {Copin}, {Corcione}, {Costille}, {Courbin}, {Degaudenzi}, {Douspis}, {Dubath}, {Duncan}, {Dusini}, {Farrens}, {Ferriol}, {Fosalba}, {Fourmanoit}, {Frailis}, {Franceschi}, {Franzetti}, {Fumana}, {Gillis}, {Giocoli}, {Grazian}, {Grupp}, {Haugan}, {Holmes}, {Hormuth}, {Hudelot}, {Kermiche}, {Kiessling}, {Kilbinger}, {Kohley}, {Kubik}, {K{\"u}mmel}, {Kunz}, {Kurki-Suonio}, {Lahav}, {Ligori}, {Lilje}, {Lloro}, {Mansutti}, {Marggraf}, {Markovic}, {Marulli}, {Massey}, {Maurogordato}, {Melchior}, {Merlin}, {Meylan}, {Mohr}, {Moresco}, {Morin}, {Moscardini}, {Munari}, {Nichol}, {Padilla}, {Paltani}, {Peacock}, {Pedersen}, {Pettorino}, {Pires}, {Poncet}, {Popa}, {Pozzetti}, {Raison}, {Rebolo}, {Rhodes}, {Rix}, {Roncarelli}, {Rossetti}, {Saglia}, {Schneider}, {Schrabback}, {Secroun}, {Seidel}, {Serrano}, {Sirignano}, {Sirri}, {Skottfelt}, {Stanco}, {Starck}, {Tallada-Cresp{\'\i}}, {Tavagnacco}, {Taylor}, {Teplitz}, {Toledo-Moreo}, {Torradeflot}, {Trifoglio}, {Valentijn}, {Valenziano}, {Verdoes Kleijn}, {Wang}, {Welikala}, {Weller}, {Wetzstein}, {Zamorani}, {Zoubian}, {Andreon}, {Baldi}, {Bardelli}, {Boucaud}, {Camera}, {Di Ferdinando}, {Fabbian}, {Farinelli}, {Galeotta}, {Graci{\'a}-Carpio}, {Maino}, {Medinaceli}, {Mei}, {Neissner}, {Polenta}, {Renzi}, {Romelli}, {Rosset}, {Sureau}, {Tenti}, {Vassallo}, {Zucca}, {Baccigalupi}, {Balaguera-Antol{\'\i}nez}, {Battaglia}, {Biviano}, {Borgani}, {Bozzo}, {Cabanac}, {Cappi}, {Casas}, {Castignani}, {Colodro-Conde}, {Coupon}, {Courtois}, {Cuby}, {de la Torre}, {Desai}, {Dole}, {Fabricius}, {Farina}, {Ferreira}, {Finelli}, {Flose-Reimberg}, {Fotopoulou}, {Ganga}, {Gozaliasl}, {Hook}, {Keihanen}, {Kirkpatrick}, {Liebing}, {Lindholm}, {Mainetti}, {Martinelli}, {Martinet}, {Maturi}, {McCracken}, {Metcalf}, {Morgante}, {Nightingale}, {Nucita}, {Patrizii}, {Potter}, {Riccio}, {S{\'a}nchez}, {Sapone}, {Schewtschenko}, {Schultheis}, {Scottez}, {Teyssier}, {Tutusaus}, {Valiviita}, {Viel}, {Vriend}, \& {Whittaker}}]{Euclid}
{Euclid Collaboration}, {Scaramella}, R., {Amiaux}, J., {et~al.} 2022, \aap, 662, A112

\bibitem[{Fagioli {et~al.}(2016)Fagioli, Carollo, Renzini, Lilly, Onodera, \& Tacchella}]{fagioliMinorMergersProgenitor2016}
Fagioli, M., Carollo, C.~M., Renzini, A., {et~al.} 2016, \apj, 831, 173

\bibitem[{Fan {et~al.}(2008)Fan, Lapi, De~Zotti, \& Danese}]{fan_dramatic_2008}
Fan, L., Lapi, A., De~Zotti, G., \& Danese, L. 2008, \apj, 689, L101

\bibitem[{{Gardner} {et~al.}(2023){Gardner}, {Mather}, {Abbott}, {Abell}, {Abernathy}, {Abney}, {Abraham}, {Abraham}, {Abul-Huda}, {Acton}, {Adams}, {Adams}, {Adler}, {Adriaensen}, {Aguilar}, {Ahmed}, {Ahmed}, {Ahmed}, {Albat}, {Albert}, {Alberts}, {Aldridge}, {Allen}, {Allen}, {Altenburg}, {Altunc}, {Alvarez}, {{\'A}lvarez-M{\'a}rquez}, {Alves de Oliveira}, {Ambrose}, {Anandakrishnan}, {Andersen}, {Anderson}, {Anderson}, {Anderson}, {Anderson}, {Aprea}, {Archer}, {Arenberg}, {Argyriou}, {Arribas}, {Artigau}, {Arvai}, {Atcheson}, {Atkinson}, {Averbukh}, {Aymergen}, {Bacinski}, {Baggett}, {Bagnasco}, {Baker}, {Balzano}, {Banks}, {Baran}, {Barker}, {Barrett}, {Barringer}, {Barto}, {Bast}, {Baudoz}, {Baum}, {Beatty}, {Beaulieu}, {Bechtold}, {Beck}, {Beddard}, {Beichman}, {Bellagama}, {Bely}, {Berger}, {Bergeron}, {Bernier}, {Bertch}, {Beskow}, {Betz}, {Biagetti}, {Birkmann}, {Bjorklund}, {Blackwood}, {Blazek}, {Blossfeld}, {Bluth}, {Boccaletti}, {Boegner}, {Bohlin}, {Boia}, {B{\"o}ker}, {Bonaventura}, {Bond}, {Bosley}, {Boucarut}, {Bouchet}, {Bouwman}, {Bower}, {Bowers}, {Bowers}, {Boyce}, {Boyer}, {Boyer}, {Boyer}, {Boyer}, {Bradley}, {Brady}, {Brandl}, {Brannen}, {Breda}, {Bremmer}, {Brennan}, {Bresnahan}, {Bright}, {Broiles}, {Bromenschenkel}, {Brooks}, {Brooks}, {Brown}, {Brown}, {Brown}, {Bruce}, {Bryson}, {Bujanda}, {Bullock}, {Bunker}, {Bureo}, {Burt}, {Bush}, {Bushouse}, {Bussman}, {Cabaud}, {Cale}, {Calhoon}, {Calvani}, {Canipe}, {Caputo}, {Cara}, {Carey}, {Case}, {Cesari}, {Cetorelli}, {Chance}, {Chandler}, {Chaney}, {Chapman}, {Charlot}, {Chayer}, {Cheezum}, {Chen}, {Chen}, {Cherinka}, {Chichester}, {Chilton}, {Chittiraibalan}, {Clampin}, {Clark}, {Clark}, {Clark}, {Claybrooks}, {Cleveland}, {Cohen}, {Cohen}, {Col{\'o}n}, {Coleman}, {Colina}, {Comber}, {Comeau}, {Comer}, {Conde Reis}, {Connolly}, {Conroy}, {Contos}, {Contreras}, {Cook}, {Cooper}, {Cooper}, {Correia}, {Correnti}, {Cossou}, {Costanza}, {Coulais}, {Cox}, {Coyle}, {Cracraft}, {Crew}, {Curtis}, {Cusveller}, {Da Costa Maciel}, {Dailey}, {Daugeron}, {Davidson}, {Davies}, {Davis}, {Davis}, {Day}, {de Chambure}, {de Jong}, {De Marchi}, {Dean}, {Decker}, {Delisa}, {Dell}, {Dellagatta}, {Dembinska}, {Demosthenes}, {Dencheva}, {Deneu}, {DePriest}, {Deschenes}, {Dethienne}, {Detre}, {Diaz}, {Dicken}, {DiFelice}, {Dillman}, {Disharoon}, {Dixon}, {Doggett}, {Dominguez}, {Donaldson}, {Doria-Warner}, {Santos}, {Doty}, {Douglas}, {Doyon}, {Dressler}, {Driggers}, {Driggers}, {Dunn}, {DuPrie}, {Dupuis}, {Durning}, {Dutta}, {Earl}, {Eccleston}, {Ecobichon}, {Egami}, {Ehrenwinkler}, {Eisenhamer}, {Eisenhower}, {Eisenstein}, {El Hamel}, {Elie}, {Elliott}, {Elliott}, {Engesser}, {Espinoza}, {Etienne}, {Etxaluze}, {Evans}, {Fabreguettes}, {Falcolini}, {Falini}, {Fatig}, {Feeney}, {Feinberg}, {Fels}, {Ferdous}, {Ferguson}, {Ferrarese}, {Ferreira}, {Ferruit}, {Ferry}, {Filippazzo}, {Firre}, {Fix}, {Flagey}, {Flanagan}, {Fleming}, {Florian}, {Flynn}, {Foiadelli}, {Fontaine}, {Fontanella}, {Forshay}, {Fortner}, {Fox}, {Framarini}, {Francisco}, {Franck}, {Franx}, {Franz}, {Friedman}, {Friend}, {Frost}, {Fu}, {Fullerton}, {Gaillard}, {Galkin}, {Gallagher}, {Galyer}, {Garc{\'\i}a Mar{\'\i}n}, {Gardner}, {Garland}, {Garrett}, {Gasman}, {G{\'a}sp{\'a}r}, {Gastaud}, {Gaudreau}, {Gauthier}, {Geers}, {Geithner}, {Gennaro}, {Gerber}, {Gereau}, {Giampaoli}, {Giardino}, {Gibbons}, {Gilbert}, {Gilman}, {Girard}, {Giuliano}, {Gkountis}, {Glasse}, {Glassmire}, {Glauser}, {Glazer}, {Goldberg}, {Golimowski}, {Gonzaga}, {Gordon}, {Gordon}, {Goudfrooij}, {Gough}, {Graham}, {Grau}, {Green}, {Greene}, {Greene}, {Greenfield}, {Greenhouse}, {Greve}, {Greville}, {Grimaldi}, {Groe}, {Groebner}, {Grumm}, {Grundy}, {G{\"u}del}, {Guillard}, {Guldalian}, {Gunn}, {Gurule}, {Gutman}, {Guy}, {Guyot}, {Hack}, {Haderlein}, {Hagan}, {Hagedorn}, {Hainline}, {Haley}, {Hami}, {Hamilton}, {Hammann}, {Hammel}, {Hanley}, {Hansen}, {Hardy}, {Harnisch}, {Harr}, {Harris}, {Hart}, {Hartig}, {Hasan}, {Hashim}, {Hashimoto}, {Haskins}, {Hawkins}, {Hayden}, {Hayden}, {Healy}, {Hecht}, {Heeg}, {Hejal}, {Helm}, {Hengemihle}, {Henning}, {Henry}, {Henry}, {Henshaw}, {Hernandez}, {Herrington}, {Heske}, {Hesman}, {Hickey}, {Hilbert}, {Hines}, {Hinz}, {Hirsch}, {Hitcho}, {Hodapp}, {Hodge}, {Hoffman}, {Holfeltz}, {Holler}, {Hoppa}, {Horner}, {Howard}, {Howard}, {Huber}, {Hunkeler}, {Hunter}, {Hunter}, {Hurd}, {Hurst}, {Hutchings}, {Hylan}, {Ignat}, {Illingworth}, {Irish}, {Isaacs}, {Jackson}, {Jaffe}, {Jahic}, {Jahromi}, {Jakobsen}, {James}, {James}, {James}, {Jamieson}, {Jandra}, {Jayawardhana}, {Jedrzejewski}, {Jeffers}, {Jensen}, {Joanne}, {Johns}, {Johnson}, {Johnson}, {Johnson}, {Johnson}, {Johnson}, {Johnson}, {Johnstone}, {Jollet}, {Jones}, {Jones}, {Jones}, {Jones}, {Jones}, {Jordan}, {Jordan}, {Jue}, {Jurkowski}, {Justis}, {Justtanont}, {Kaleida}, {Kalirai}, {Kalmanson}, {Kaltenegger}, {Kammerer}, {Kan}, {Kanarek}, {Kao}, {Karakla}, {Karl}, {Kassin}, {Kauffman}, {Kavanagh}, {Kelley}, {Kelly}, {Kendrew}, {Kennedy}, {Kenny}, {Keski-Kuha}, {Keyes}, {Khan}, {Kidwell}, {Kimble}, {King}, {King}, {Kinzel}, {Kirk}, {Kirkpatrick}, {Klaassen}, {Klingemann}, {Klintworth}, {Knapp}, {Knight}, {Knollenberg}, {Knutsen}, {Koehler}, {Koekemoer}, {Kofler}, {Kontson}, {Kovacs}, {Kozhurina-Platais}, {Krause}, {Kriss}, {Krist}, {Kristoffersen}, {Krogel}, {Krueger}, {Kulp}, {Kumari}, {Kwan}, {Kyprianou}, {Labador}, {Labiano}, {Lafreni{\`e}re}, {Lagage}, {Laidler}, {Laine}, {Laird}, {Lajoie}, {Lallo}, {Lam}, {LaMassa}, {Lambros}, {Lampenfield}, {Lander}, {Langston}, {Larson}, {Larson}, {LaVerghetta}, {Law}, {Lawrence}, {Lee}, {Lee}, {Lee}, {Leisenring}, {Leveille}, {Levenson}, {Levi}, {Levine}, {Lewis}, {Lewis}, {Lewis}, {Libralato}, {Lidon}, {Liebrecht}, {Lightsey}, {Lilly}, {Lim}, {Lim}, {Ling}, {Link}, {Link}, {Lipinski}, {Liu}, {Lo}, {Lobmeyer}, {Logue}, {Long}, {Long}, {Long}, {Long}, {L{\'o}pez-Caniego}, {Lotz}, {Love-Pruitt}, {Lubskiy}, {Luers}, {Luetgens}, {Luevano}, {Lui}, {Lund}, {Lundquist}, {Lunine}, {L{\"u}tzgendorf}, {Lynch}, {MacDonald}, {MacDonald}, {Macias}, {Macklis}, {Maghami}, {Maharaja}, {Maiolino}, {Makrygiannis}, {Malla}, {Malumuth}, {Manjavacas}, {Marini}, {Marrione}, {Marston}, {Martel}, {Martin}, {Martin}, {Martinez}, {Maschmann}, {Masci}, {Masetti}, {Maszkiewicz}, {Matthews}, {Matuskey}, {McBrayer}, {McCarthy}, {McCaughrean}, {McClare}, {McClare}, {McCloskey}, {McClurg}, {McCoy}, {McElwain}, {McGregor}, {McGuffey}, {McKay}, {McKenzie}, {McLean}, {McMaster}, {McNeil}, {De Meester}, {Mehalick}, {Meixner}, {Mel{\'e}ndez}, {Menzel}, {Menzel}, {Merz}, {Mesterharm}, {Meyer}, {Meyett}, {Meza}, {Midwinter}, {Milam}, {Miller}, {Miller}, {Miskey}, {Misselt}, {Mitchell}, {Mohan}, {Montoya}, {Moran}, {Morishita}, {Moro-Mart{\'\i}n}, {Morrison}, {Morrison}, {Morse}, {Moschos}, {Moseley}, {Mosier}, {Mosner}, {Mountain}, {Muckenthaler}, {Mueller}, {Mueller}, {Muhiem}, {M{\"u}hlmann}, {Mullally}, {Mullen}, {Munger}, {Murphy}, {Murray}, {Muzerolle}, {Mycroft}, {Myers}, {Myers}, {Myers}, {Myers}, {Myrick}, {Nagle}, {Nayak}, {Naylor}, {Neff}, {Nelan}, {Nella}, {Nguyen}, {Nguyen}, {Nickson}, {Nidhiry}, {Niedner}, {Nieto-Santisteban}, {Nikolov}, {Nishisaka}, {Noriega-Crespo}, {Nota}, {O'Mara}, {Oboryshko}, {O'Brien}, {Ochs}, {Offenberg}, {Ogle}, {Ohl}, {Olmsted}, {Osborne}, {O'Shaughnessy}, {{\"O}stlin}, {O'Sullivan}, {Otor}, {Ottens}, {Ouellette}, {Outlaw}, {Owens}, {Pacifici}, {Page}, {Paranilam}, {Park}, {Parrish}, {Paschal}, {Patapis}, {Patel}, {Patrick}, {Pattishall}, {Paul}, {Paul}, {Pauly}, {Pavlovsky}, {Pe{\~n}a-Guerrero}, {Pedder}, {Peek}, {Pelham}, {Penanen}, {Perriello}, {Perrin}, {Perrine}, {Perrygo}, {Peslier}, {Petach}, {Peterson}, {Pfarr}, {Pierson}, {Pietraszkiewicz}, {Pilchen}, {Pipher}, {Pirzkal}, {Pitman}, {Player}, {Plesha}, {Plitzke}, {Pohner}, {Poletis}, {Pollizzi}, {Polster}, {Pontius}, {Pontoppidan}, {Porges}, {Potter}, {Prescott}, {Proffitt}, {Pueyo}, {Quispe Neira}, {Radich}, {Rager}, {Rameau}, {Ramey}, {Ramos Alarcon}, {Rampini}, {Rapp}, {Rashford}, {Rauscher}, {Ravindranath}, {Rawle}, {Rawlings}, {Ray}, {Regan}, {Rehm}, {Rehm}, {Reid}, {Reis}, {Renk}, {Reoch}, {Ressler}, {Rest}, {Reynolds}, {Richon}, {Richon}, {Ridgaway}, {Riedel}, {Rieke}, {Rieke}, {Rifelli}, {Rigby}, {Riggs}, {Ringel}, {Ritchie}, {Rix}, {Robberto}, {Robinson}, {Robinson}, {Robinson}, {Rock}, {Rodriguez}, {Rodr{\'\i}guez del Pino}, {Roellig}, {Rohrbach}, {Roman}, {Romelfanger}, {Romo}, {Rosales}, {Rose}, {Roteliuk}, {Roth}, {Rothwell}, {Rouzaud}, {Rowe}, {Rowlands}, {Roy}, {Royer}, {Rui}, {Rumler}, {Rumpl}, {Russ}, {Ryan}, {Ryan}, {Saad}, {Sabata}, {Sabatino}, {Sabbi}, {Sabelhaus}, {Sabia}, {Sahu}, {Saif}, {Salvignol}, {Samara-Ratna}, {Samuelson}, {Sanders}, {Sappington}, {Sargent}, {Sauer}, {Savadkin}, {Sawicki}, {Schappell}, {Scheffer}, {Scheithauer}, {Scherer}, {Schiff}, {Schlawin}, {Schmeitzky}, {Schmitz}, {Schmude}, {Schneider}, {Schreiber}, {Schroeven-Deceuninck}, {Schultz}, {Schwab}, {Schwartz}, {Scoccimarro}, {Scott}, {Scott}, {Seaton}, {Seely}, {Seery}, {Seidleck}, {Sembach}, {Shanahan}, {Shaughnessy}, {Shaw}, {Shay}, {Sheehan}, {Sheth}, {Shih}, {Shivaei}, {Siegel}, {Sienkiewicz}, {Simmons}, {Simon}, {Sirianni}, {Sivaramakrishnan}, {Slade}, {Sloan}, {Slocum}, {Slowinski}, {Smith}, {Smith}, {Smith}, {Smith}, {Smith}, {Smith}, {Smolik}, {Soderblom}, {Sohn}, {Sokol}, {Sonneborn}, {Sontag}, {Sooy}, {Soummer}, {Southwood}, {Spain}, {Sparmo}, {Speer}, {Spencer}, {Sprofera}, {Stallcup}, {Stanley}, {Stansberry}, {Stark}, {Starr}, {Stassi}, {Steck}, {Steeley}, {Stephens}, {Stephenson}, {Stewart}, {Stiavelli}, {}, {Strada}, {Straughn}, {Streetman}, {Strickland}, {Strobele}, {Stuhlinger}, {Stys}, {Such}, {Sukhatme}, {Sullivan}, {Sullivan}, {Sumner}, {Sun}, {Sunnquist}, {Swade}, {Swam}, {Swenton}, {Swoish}, {Tam Litten}, {Tamas}, {Tao}, {Taylor}, {Taylor}, {te Plate}, {Van Tea}, {Teague}, {Telfer}, {Temim}, {Texter}, {Thatte}, {Thompson}, {Thompson}, {Thomson}, {Thronson}, {Tierney}, {Tikkanen}, {Tinnin}, {Tippet}, {Todd}, {Tran},
  {Trauger}, {Trejo}, {Vinh Truong}, {Tsukamoto}, {Tufail}, {Tumlinson}, {Tustain}, {Tyra}, {Ubeda}, {Underwood}, {Uzzo}, {Vaclavik}, {Valenduc}, {Valenti}, {Van Campen}, {van de Wetering}, {Van Der Marel}, {van Haarlem}, {Vandenbussche}, {van Dishoeck}, {Vanterpool}, {Vernoy}, {Vila Costas}, {Volk}, {Voorzaat}, {Voyton}, {Vydra}, {Waddy}, {Waelkens}, {Wahlgren}, {Walker}, {Wander}, {Warfield}, {Warner}, {Wasiak}, {Wasiak}, {Wehner}, {Weiler}, {Weilert}, {Weiss}, {Wells}, {Welty}, {Wheate}, {Wheeler}, {White}, {Whitehouse}, {Whiteleather}, {Whitman}, {Williams}, {Willmer}, {Willott}, {Willoughby}, {Wilson}, {Wilson}, {Wilson}, {Windhorst}, {Wislowski}, {Wolfe}, {Wolfe}, {Wolff}, {Wondel}, {Woo}, {Woods}, {Worden}, {Workman}, {Wright}, {Wu}, {Wu}, {Wun}, {Wymer}, {Yadetie}, {Yan}, {Yang}, {Yates}, {Yeager}, {Yerger}, {Young}, {Young}, {Yu}, {Yu}, {Zak}, {Zeidler}, {Zepp}, {Zhou}, {Zincke}, {Zonak}, \& {Zondag}}]{Gardner_JWST}
{Gardner}, J.~P., {Mather}, J.~C., {Abbott}, R., {et~al.} 2023, \pasp, 135, 068001

\bibitem[{Hamadouche {et~al.}(2022)Hamadouche, Carnall, McLure, Dunlop, McLeod, Cullen, Begley, Bolzonella, Buitrago, Castellano, Cucciati, Fontana, Gargiulo, Moresco, Pozzetti, \& Zamorani}]{hamadouche2022}
Hamadouche, M.~L., Carnall, A.~C., McLure, R.~J., {et~al.} 2022, \mnras, 512, 1262

\bibitem[{Hilz {et~al.}(2013)Hilz, Naab, \& Ostriker}]{hilz_how_2013}
Hilz, M., Naab, T., \& Ostriker, J.~P. 2013, \mnras, 429, 2924

\bibitem[{{Hopkins} {et~al.}(2013){Hopkins}, {Driver}, {Brough}, {Owers}, {Bauer}, {Gunawardhana}, {Cluver}, {Colless}, {Foster}, {Lara-L{\'o}pez}, {Roseboom}, {Sharp}, {Steele}, {Thomas}, {Baldry}, {Brown}, {Liske}, {Norberg}, {Robotham}, {Bamford}, {Bland-Hawthorn}, {Drinkwater}, {Loveday}, {Meyer}, {Peacock}, {Tuffs}, {Agius}, {Alpaslan}, {Andrae}, {Cameron}, {Cole}, {Ching}, {Christodoulou}, {Conselice}, {Croom}, {Cross}, {De Propris}, {Delhaize}, {Dunne}, {Eales}, {Ellis}, {Frenk}, {Graham}, {Grootes}, {H{\"a}u{\ss}ler}, {Heymans}, {Hill}, {Hoyle}, {Hudson}, {Jarvis}, {Johansson}, {Jones}, {van Kampen}, {Kelvin}, {Kuijken}, {L{\'o}pez-S{\'a}nchez}, {Maddox}, {Madore}, {Maraston}, {McNaught-Roberts}, {Nichol}, {Oliver}, {Parkinson}, {Penny}, {Phillipps}, {Pimbblet}, {Ponman}, {Popescu}, {Prescott}, {Proctor}, {Sadler}, {Sansom}, {Seibert}, {Staveley-Smith}, {Sutherland}, {Taylor}, {Van Waerbeke}, {V{\'a}zquez-Mata}, {Warren}, {Wijesinghe}, {Wild}, \& {Wilkins}}]{GAMA2}
{Hopkins}, A.~M., {Driver}, S.~P., {Brough}, S., {et~al.} 2013, \mnras, 430, 2047

\bibitem[{{Hopkins} {et~al.}(2010){Hopkins}, {Bundy}, {Hernquist}, {Wuyts}, \& {Cox}}]{hopkins2010}
{Hopkins}, P.~F., {Bundy}, K., {Hernquist}, L., {Wuyts}, S., \& {Cox}, T.~J. 2010, \mnras, 401, 1099

\bibitem[{{Hopkins} {et~al.}(2009){Hopkins}, {Hernquist}, {Cox}, {Keres}, \& {Wuyts}}]{hopkins2009}
{Hopkins}, P.~F., {Hernquist}, L., {Cox}, T.~J., {Keres}, D., \& {Wuyts}, S. 2009, \apj, 691, 1424

\bibitem[{{Huang} {et~al.}(2018){Huang}, {Leauthaud}, {Greene}, {Bundy}, {Lin}, {Tanaka}, {Miyazaki}, \& {Komiyama}}]{Huang2018}
{Huang}, S., {Leauthaud}, A., {Greene}, J.~E., {et~al.} 2018, \mnras, 475, 3348

\bibitem[{{Huang} {et~al.}(2020){Huang}, {Leauthaud}, {Hearin}, {Behroozi}, {Bradshaw}, {Ardila}, {Speagle}, {Tenneti}, {Bundy}, {Greene}, {Sif{\'o}n}, \& {Bahcall}}]{Huang2020}
{Huang}, S., {Leauthaud}, A., {Hearin}, A., {et~al.} 2020, \mnras, 492, 3685

\bibitem[{{Kauffmann} {et~al.}(2003){Kauffmann}, {Heckman}, {White}, {Charlot}, {Tremonti}, {Brinchmann}, {Bruzual}, {Peng}, {Seibert}, {Bernardi}, {Blanton}, {Brinkmann}, {Castander}, {Cs{\'a}bai}, {Fukugita}, {Ivezic}, {Munn}, {Nichol}, {Padmanabhan}, {Thakar}, {Weinberg}, \& {York}}]{Kauffmann2003}
{Kauffmann}, G., {Heckman}, T.~M., {White}, S. D.~M., {et~al.} 2003, \mnras, 341, 33

\bibitem[{{Kawinwanichakij} {et~al.}(2020){Kawinwanichakij}, {Papovich}, {Ciardullo}, {Finkelstein}, {Stevans}, {Wold}, {Jogee}, {Sherman}, {Florez}, \& {Gronwall}}]{Kawinwanichakij2020}
{Kawinwanichakij}, L., {Papovich}, C., {Ciardullo}, R., {et~al.} 2020, \apj, 892, 7

\bibitem[{Kuijken {et~al.}(2019)Kuijken, Heymans, Dvornik, Hildebrandt, De~Jong, Wright, Erben, Bilicki, Giblin, Shan, Getman, Grado, Hoekstra, Miller, Napolitano, Paolilo, Radovich, Schneider, Sutherland, Tewes, Tortora, Valentijn, \& Verdoes~Kleijn}]{kuijken_fourth_2019}
Kuijken, K., Heymans, C., Dvornik, A., {et~al.} 2019, \aap, 625, A2

\bibitem[{Li {et~al.}(2022)Li, Napolitano, Roy, Tortora, La~Barbera, Sonnenfeld, Qiu, \& Liu}]{GaLNet2022}
Li, R., Napolitano, N.~R., Roy, N., {et~al.} 2022, \apj, 929, 152

\bibitem[{{Moster} {et~al.}(2018){Moster}, {Naab}, \& {White}}]{2018moster}
{Moster}, B.~P., {Naab}, T., \& {White}, S. D.~M. 2018, \mnras, 477, 1822

\bibitem[{Naab {et~al.}(2009)Naab, Johansson, \& Ostriker}]{naab_minor_2009}
Naab, T., Johansson, P.~H., \& Ostriker, J.~P. 2009, \apj, 699, L178, publisher: The American Astronomical Society

\bibitem[{{Navarro} {et~al.}(1996){Navarro}, {Frenk}, \& {White}}]{NFW}
{Navarro}, J.~F., {Frenk}, C.~S., \& {White}, S. D.~M. 1996, \apj, 462, 563

\bibitem[{Newman {et~al.}(2012)Newman, Ellis, Bundy, \& Treu}]{newman2012}
Newman, A.~B., Ellis, R.~S., Bundy, K., \& Treu, T. 2012, \apj, 746, 162

\bibitem[{{Nipoti}(2025)}]{Nipoti2025}
{Nipoti}, C. 2025, \aap, 697, A74

\bibitem[{{Nipoti} {et~al.}(2009{\natexlab{a}}){Nipoti}, {Treu}, {Auger}, \& {Bolton}}]{nipotitreu09}
{Nipoti}, C., {Treu}, T., {Auger}, M.~W., \& {Bolton}, A.~S. 2009{\natexlab{a}}, \apjl, 706, L86

\bibitem[{{Nipoti} {et~al.}(2009{\natexlab{b}}){Nipoti}, Treu, \& Bolton}]{nipoti2009}
{Nipoti}, C., Treu, T., \& Bolton, A.~S. 2009{\natexlab{b}}, \apj, 703, 1531

\bibitem[{{Nipoti} {et~al.}(2012){Nipoti}, {Treu}, {Leauthaud}, {Bundy}, {Newman}, \& {Auger}}]{nipoti12}
{Nipoti}, C., {Treu}, T., {Leauthaud}, A., {et~al.} 2012, \mnras, 422, 1714

\bibitem[{Oser {et~al.}(2011)Oser, Naab, Ostriker, \& Johansson}]{oser_cosmological_2011}
Oser, L., Naab, T., Ostriker, J.~P., \& Johansson, P.~H. 2011, \apj, 744, 63, publisher: The American Astronomical Society

\bibitem[{{Paulino-Afonso} {et~al.}(2022){Paulino-Afonso}, {Gonz{\'a}lez-Gait{\'a}n}, {Galbany}, {Maria Mour{\~a}o}, {Angus}, {Smith}, {Anderson}, {Lyman}, {Kuncarayakti}, \& {Rodrigues}}]{Paulino2022}
{Paulino-Afonso}, A., {Gonz{\'a}lez-Gait{\'a}n}, S., {Galbany}, L., {et~al.} 2022, \aap, 662, A86

\bibitem[{{Roy} {et~al.}(2018){Roy}, {Napolitano}, {La Barbera}, {Tortora}, {Getman}, {Radovich}, {Capaccioli}, {Brescia}, {Cavuoti}, {Longo}, {Raj}, {Puddu}, {Covone}, {Amaro}, {Vellucci}, {Grado}, {Kuijken}, {Verdoes Kleijn}, \& {Valentijn}}]{KiDS_Roy}
{Roy}, N., {Napolitano}, N.~R., {La Barbera}, F., {et~al.} 2018, \mnras, 480, 1057

\bibitem[{{Saglia} {et~al.}(2010){Saglia}, {S{\'a}nchez-Bl{\'a}zquez}, {Bender}, {Simard}, {Desai}, {Arag{\'o}n-Salamanca}, {Milvang-Jensen}, {Halliday}, {Jablonka}, {Noll}, {Poggianti}, {Clowe}, {De Lucia}, {Pell{\'o}}, {Rudnick}, {Valentinuzzi}, {White}, \& {Zaritsky}}]{saglia2010}
{Saglia}, R.~P., {S{\'a}nchez-Bl{\'a}zquez}, P., {Bender}, R., {et~al.} 2010, \aap, 524, A6

\bibitem[{{Sersic}(1968)}]{Sersic1968}
{Sersic}, J.~L. 1968, {Atlas de Galaxias Australes}

\bibitem[{{Sonnenfeld}(2020)}]{Alessandro20}
{Sonnenfeld}, A. 2020, \aap, 641, A143

\bibitem[{Sonnenfeld {et~al.}(2014)Sonnenfeld, Nipoti, \& Treu}]{sonnenfeld2014}
Sonnenfeld, A., Nipoti, C., \& Treu, T. 2014, \apj, 786, 89

\bibitem[{{Spavone} {et~al.}(2021){Spavone}, {Krajnovi{\'c}}, {Emsellem}, {Iodice}, \& {den Brok}}]{spavone2021}
{Spavone}, M., {Krajnovi{\'c}}, D., {Emsellem}, E., {Iodice}, E., \& {den Brok}, M. 2021, \aap, 649, A161

\bibitem[{{Suess} {et~al.}(2019{\natexlab{a}}){Suess}, {Kriek}, {Price}, \& {Barro}}]{Suess2019a}
{Suess}, K.~A., {Kriek}, M., {Price}, S.~H., \& {Barro}, G. 2019{\natexlab{a}}, \apj, 877, 103

\bibitem[{{Suess} {et~al.}(2019{\natexlab{b}}){Suess}, {Kriek}, {Price}, \& {Barro}}]{Suess2019b}
{Suess}, K.~A., {Kriek}, M., {Price}, S.~H., \& {Barro}, G. 2019{\natexlab{b}}, \apjl, 885, L22

\bibitem[{{Suess} {et~al.}(2020){Suess}, {Kriek}, {Price}, \& {Barro}}]{Suess2020}
{Suess}, K.~A., {Kriek}, M., {Price}, S.~H., \& {Barro}, G. 2020, \apjl, 899, L26

\bibitem[{{Suess} {et~al.}(2023){Suess}, {Williams}, {Robertson}, {Ji}, {Johnson}, {Nelson}, {Alberts}, {Hainline}, {D'Eugenio}, {{\"U}bler}, {Rieke}, {Rieke}, {Bunker}, {Carniani}, {Charlot}, {Eisenstein}, {Maiolino}, {Stark}, {Tacchella}, \& {Willott}}]{Suess2023}
{Suess}, K.~A., {Williams}, C.~C., {Robertson}, B., {et~al.} 2023, \apjl, 956, L42

\bibitem[{{Tal} \& {van Dokkum}(2011)}]{tal_2011_faint}
{Tal}, T. \& {van Dokkum}, P.~G. 2011, \apj, 731, 89

\bibitem[{{Taylor} {et~al.}(2011){Taylor}, {Hopkins}, {Baldry}, {Brown}, {Driver}, {Kelvin}, {Hill}, {Robotham}, {Bland-Hawthorn}, {Jones}, {Sharp}, {Thomas}, {Liske}, {Loveday}, {Norberg}, {Peacock}, {Bamford}, {Brough}, {Colless}, {Cameron}, {Conselice}, {Croom}, {Frenk}, {Gunawardhana}, {Kuijken}, {Nichol}, {Parkinson}, {Phillipps}, {Pimbblet}, {Popescu}, {Prescott}, {Sutherland}, {Tuffs}, {van Kampen}, \& {Wijesinghe}}]{Taylor2011}
{Taylor}, E.~N., {Hopkins}, A.~M., {Baldry}, I.~K., {et~al.} 2011, \mnras, 418, 1587

\bibitem[{Toft {et~al.}(2007)Toft, {van Dokkum}, Franx, Labbe, F{\"o}rster~Schreiber, Wuyts, Webb, Rudnick, Zirm, Kriek, {van der Werf}, Blakeslee, Illingworth, Rix, Papovich, \& Moorwood}]{toft2007}
Toft, S., {van Dokkum}, P., Franx, M., {et~al.} 2007, \apj, 671, 285

\bibitem[{{Tremaine} {et~al.}(1994){Tremaine}, {Richstone}, {Byun}, {Dressler}, {Faber}, {Grillmair}, {Kormendy}, \& {Lauer}}]{Tremaine94}
{Tremaine}, S., {Richstone}, D.~O., {Byun}, Y.-I., {et~al.} 1994, \aj, 107, 634

\bibitem[{Trujillo {et~al.}(2007)Trujillo, Conselice, Bundy, Cooper, Eisenhardt, \& Ellis}]{trujillo2007}
Trujillo, I., Conselice, C.~J., Bundy, K., {et~al.} 2007, \mnras, 382, 109

\bibitem[{Trujillo {et~al.}(2006)Trujillo, Feulner, Goranova, Hopp, Longhetti, Saracco, Bender, Braito, Della~Ceca, Drory, Mannucci, \& Severgnini}]{trujillo2006}
Trujillo, I., Feulner, G., Goranova, Y., {et~al.} 2006, \mnras, 373, L36

\bibitem[{{van der Wel} {et~al.}(2014){van der Wel}, Franx, {van Dokkum}, Skelton, Momcheva, Whitaker, Brammer, Bell, Rix, Wuyts, Ferguson, Holden, Barro, Koekemoer, Chang, McGrath, H{\"a}ussler, Dekel, Behroozi, Fumagalli, Leja, Lundgren, Maseda, Nelson, Wake, Patel, Labb{\'e}, Faber, Grogin, \& Kocevski}]{vanderwel3DHSTCANDELSEvolution2014}
{van der Wel}, A., Franx, M., {van Dokkum}, P.~G., {et~al.} 2014, \apj, 788, 28

\bibitem[{{van Dokkum} \& {Brammer}(2010)}]{van_dokkum_2010_hubble}
{van Dokkum}, P.~G. \& {Brammer}, G. 2010, \apjl, 718, L73

\bibitem[{{van Dokkum} \& Franx(2001)}]{vandokkumMorphologicalEvolutionAges2001}
{van Dokkum}, P.~G. \& Franx, M. 2001, \apj, 553, 90

\bibitem[{{van Dokkum} {et~al.}(2008){van Dokkum}, Franx, Kriek, Holden, Illingworth, Magee, Bouwens, Marchesini, Quadri, Rudnick, Taylor, \& Toft}]{vandokkum2008}
{van Dokkum}, P.~G., Franx, M., Kriek, M., {et~al.} 2008, \apj, 677, L5

\bibitem[{van Dokkum {et~al.}(2010)van Dokkum, Whitaker, Brammer, Franx, Kriek, Labbé, Marchesini, Quadri, Bezanson, Illingworth, Muzzin, Rudnick, Tal, \& Wake}]{van_dokkum_growth_2010}
van Dokkum, P.~G., Whitaker, K.~E., Brammer, G., {et~al.} 2010, \apj, 709, 1018, aDS Bibcode: 2010ApJ...709.1018V

\bibitem[{{Williams} {et~al.}(2024){Williams}, {Damjanov}, {Sawicki}, {Souchereau}, {Chen}, {Desprez}, {George}, {Annunziatella}, \& {Gwyn}}]{Williams2024}
{Williams}, D.~J., {Damjanov}, I., {Sawicki}, M., {et~al.} 2024, arXiv e-prints, arXiv:2412.03662

\bibitem[{{Zahid} {et~al.}(2019){Zahid}, {Geller}, {Damjanov}, \& {Sohn}}]{zahid2019}
{Zahid}, H.~J., {Geller}, M.~J., {Damjanov}, I., \& {Sohn}, J. 2019, \apj, 878, 158

\bibitem[{Zhan(2021)}]{csst}
Zhan, H. 2021, Chinese Science Bulletin, 66, 1290

\end{thebibliography}
\end{document}